\documentclass[%
reprint,
amsmath,amssymb,
aps,
amsthm,
pra,
]{revtex4-1}

\usepackage{ifthen}
\newboolean{arxiv}
\setboolean{arxiv}{true}
\ifthenelse{\boolean{arxiv}}{\pdfoutput=1}{}

\usepackage{theorem} % \theorembodyfont
\usepackage{ascmac}
\usepackage{bbm}
\usepackage{bm} % bold math
\usepackage{braket}
\usepackage{dcolumn} % Align table columns on decimal point
\usepackage{enumerate}
\usepackage{enumitem}
\usepackage{framed}
\usepackage{graphicx} % Include figure files
\ifthenelse{\boolean{arxiv}}{\usepackage{hyperref}}{\usepackage[dvipdfmx]{hyperref}}
\usepackage{longtable}
\usepackage{mathtools}
\usepackage{multirow}
\usepackage{natbib}
\ifthenelse{\boolean{arxiv}}{}{\usepackage{pxjahyper}}
\usepackage{txfonts} % \coloneqq,\llbracket
\usepackage{xparse}

\usepackage{color}
\usepackage[most]{tcolorbox}

\ifthenelse{\boolean{arxiv}}{}{\mathtoolsset{showonlyrefs}}

\newcommand{\hypercolor}{blue}
\hypersetup{
  colorlinks,
  citecolor=\hypercolor,
  linkcolor=\hypercolor,
  urlcolor=\hypercolor,
  bookmarksopen,
  bookmarksopenlevel=4,
  bookmarksnumbered
}

\newboolean{figure}
\setboolean{figure}{true}
\newcommand{\InsertPDF}[1]{\iffigure\includegraphics[scale=1.0]{#1}\fi}

\usepackage{bm}

\theorembodyfont{\upshape}

\newtheorem{postulateno}{Postulate}

\newcounter{proof}

\ExplSyntaxOn
\NewDocumentEnvironment{proof}{o}
 {
  \par\medskip
  \noindent
  \textbf{Proof~}
 }
 {\QED\par\smallskip}
\ExplSyntaxOff

\newcounter{postulate}
\renewcommand{\thepostulate}{\arabic{postulate}}
\ExplSyntaxOn
\NewDocumentEnvironment{postulate}{oo}
 {
  \refstepcounter{postulate}
  \begin{postulateno}
  \textbf{\hspace{-0.5em}\IfNoValueTF{#2}{\thepostulate}{#2} ~(\IfNoValueTF{#1}{}{#1})}
 }
 {
  \end{postulateno}
 }
\ExplSyntaxOff

\newcommand{\mT}{\mathcal{T}}

\newcommand{\trho}{\tilde{\rho}}

\newcommand{\PS}{P_{\rm S}}

\newcommand{\ident}{\hat{1}}
\newcommand{\Real}{\mathbb{R}}
\newcommand{\Complex}{\mathbb{C}}

\newcommand{\QED}{\hspace*{0pt}\hfill $\blacksquare$}

\DeclareMathOperator{\Tr}{Tr}
\newcommand{\Trp}[1]{\mathop{\mathrm{Tr}_{#1}}}

\newcommand{\titlename}{\RA{Simple upper and lower bounds on the ultimate success probability for %
discriminating arbitrary finite-dimensional quantum processes}}

\newcommand{\supplementaltitle}{{\large \bf Supplemental Material for ``\titlename''} \\}
\newcommand{\supplementalauthor}[1]{\vspace*{0.5cm}#1 \\}
\newcommand{\supplementalaffiliation}[1]{\textit{#1}}
\newcommand{\citesupF}[1]{Sec.~\ref{#1} of the Supplemental Material (SM) \cite{SM}}
\newcommand{\citesup}[1]{Sec.~\ref{#1} of the SM \cite{SM}}

\renewcommand{\PS}{P}
\renewcommand{\ol}{\overline}
\newcommand{\oPS}[1]{\ol{P_#1}}
\newcommand{\lPS}{\underline{P}}
\newcommand{\opt}{\star}
\newcommand{\cE}{\mathcal{E}}
\newcommand{\summ}{\sum_{m=1}^M}
\newcommand{\sumk}{\sum_{k=1}^M}
\renewcommand{\ident}{\mathbbm{1}}
\newcommand{\I}{I}
\newcommand{\Pos}{\mathsf{Pos}}
\newcommand{\Chn}{\mathsf{Chn}}
\newcommand{\Den}{\mathsf{Den}}
\newcommand{\ot}{\otimes}
\newcommand{\V}{V}
\newcommand{\W}{W}
\newcommand{\NV}{N_\V}
\newcommand{\tN}{\tilde{N}}
\newcommand{\AD}{A}
\newcommand{\GAD}{G}
\newcommand{\tLambda}{\tilde{\Lambda}}
\renewcommand{\c}{\circ}
\newcommand{\C}{\mathsf{C}}
\newcommand{\gdis}{\raisebox{-.1em}{\includegraphics[scale=0.5]{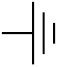}}}
\newcommand{\pc}{p_\mathrm{c}}
\newcommand{\poly}{\mathrm{poly}}
\let\ast\relax
\DeclareMathOperator{\ast}{\circledast}
\setlist[enumerate]{label=\arabic*), leftmargin=3em, itemsep=0pt, parsep=0pt, labelwidth=5em}

\definecolor{memo}{RGB}{128,0,255}
\definecolor{gray}{RGB}{128,128,128}
\definecolor{green}{RGB}{32,192,0}
\newcommand{\RA}[1]{\Red{#1}}
\newcommand{\RB}[1]{\Blue{#1}}
\newcommand{\RO}[1]{\Green{#1}}
\renewcommand{\RA}[1]{#1}
\renewcommand{\RB}[1]{#1}
\renewcommand{\RO}[1]{#1}

\newcommand{\Discard}[1]{}

\begin{document}

\preprint{APS/123-QED}

\title{\titlename}

\affiliation{%
 Quantum Information Science Research Center, Quantum ICT Research Institute, Tamagawa University,
 Machida, Tokyo 194-8610, Japan
}%

\author{Kenji Nakahira}
\affiliation{%
 Quantum Information Science Research Center, Quantum ICT Research Institute, Tamagawa University,
 Machida, Tokyo 194-8610, Japan
}%

\author{Kentaro Kato}
\affiliation{%
 Quantum Information Science Research Center, Quantum ICT Research Institute, Tamagawa University,
 Machida, Tokyo 194-8610, Japan
}%

\date{\today}

\begin{abstract}
 We consider the problem of discriminating finite-dimensional quantum processes,
 also called quantum supermaps, that can consist of multiple time steps.
 Obtaining the ultimate performance for discriminating quantum processes
 is of fundamental importance, but is challenging mainly due to
 the necessity of considering all discrimination strategies allowed by quantum mechanics,
 including entanglement-assisted strategies and adaptive strategies.
 In the case in which the processes to be discriminated have internal memories,
 the ultimate performance would generally be more difficult to analyze.
 In this Letter, we present a simple upper bound on the ultimate success probability
 for discriminating arbitrary quantum processes.
 In the special case of multi-shot channel discrimination,
 it can be shown that the ultimate success probability increases by at most a constant factor
 determined by the given channels if the number of channel evaluations increases by one.
 We also present a lower bound based on Bayesian updating,
 which has a low computational cost.
 Our numerical experiments demonstrate that the proposed bounds are reasonably tight.
 The proposed bounds do not explicitly depend on any quantum phenomena, and can be
 readily extended to a general operational probabilistic theory.
\end{abstract}

% PACS 03.67.Hk: Quantum communication
\pacs{03.67.Hk}% PACS, the Physics and Astronomy
\maketitle

%\onecolumngrid

A quantum process, which is a mathematical object that models the probabilistic behavior
of quantum devices, plays an essential role in quantum information science.
Discriminating between quantum processes is a fundamental and challenging problem,
which forms the basis of a large class of problems in quantum information theory
such as quantum communication, quantum cryptography, and quantum metrology.
The simplest instance of this problem is a quantum state discrimination problem,
which has been widely studied since the end of the 1960's \cite{Hel-1969,Hol-1973,Yue-Ken-Lax-1975}.
Since the maximum success probability is often quite difficult to obtain accurately,
its upper and lower bounds have been developed
\cite{Bel-1975,Hay-Kaw-Kob-2008,Mon-2008,Qiu-2008,Tys-2009,Qiu-Li-2010}.
Discrimination problems of quantum measurements
\cite{Ji-Fen-Dua-Yin-2006,Zim-Hei-2008,Zim-Hei-Sed-2009,Sed-Zim-2014,Puc-Paw-Kra-Kuk-2018,Kra-Paw-Puc-2020,
Dat-Bis-Sah-Aug-2020}
and quantum channels
\RO{\cite{Aci-2001,Sac-2005,Sac-2005-EB,Li-Qiu-2008,Pir-Lup-2017,Pir-Lau-Lup-Per-2019}}
are also particular instances.
In quantum channel discrimination, entanglement with an ancillary system
and an adaptive strategy may be required to achieve the ultimate performance,
which makes this problem difficult in general.
A quantum process describes the most general transformation that maps channels to channels
\cite{Chi-Dar-Per-2008-supermap,Chi-Dar-Per-2008}.
A process can consist of several memory channels
\cite{Mac-Pal-2002,Yeo-Ske-2003,Bow-Man-2004,Kre-Wer-2005,Ple-Vir-2007},
whose output states can depend on the previous input states.
\RA{As an example of process discrimination, we can consider the problem of retrieving
the value of the bit that is encoded into the reflectivity of a certain memory cell,
which is often referred to as quantum reading \cite{Pir-2011}.
This problem can be seen as the discrimination of two processes,
even when a finite number of uses of the memory cell are allowed
and the reflectivity may change depending on the previous inputs to it.}
Although finding the ultimate performance for discriminating such processes is extremely difficult,
it is of fundamental importance in various fields including
quantum cryptography \cite{Dar-Kre-Sch-Wer-2007},
quantum game theory \cite{Gut-Wat-2007}, and quantum algorithms.

In this Letter, we derive a simple upper bound on the ultimate success probability
for discriminating arbitrary finite-dimensional quantum processes.
In the special case of multi-shot channel discrimination,
our approach can ensure that the ultimate success probability increases by
at most a constant factor, which is determined by the given channels,
if the number of channel evaluations increases by one.
Note that an upper bound for channel discrimination has been reported very recently \cite{Zhu-Pir-2020},
which is based on port-based teleportation \cite{Ish-Hir-2008,Ish-Hir-2009}.
We present \RA{numerical simulations that show that, at least in a certain multi-shot channel
discrimination problem,} our upper bound is significantly tighter
than that of Ref.~\cite{Zhu-Pir-2020}.

A tight lower bound is also required to accurately evaluate the ultimate performance.
Since the success probability of any discrimination allowed by quantum mechanics
yields a lower bound on the ultimate success probability,
a natural approach to derive such a bound is to find good discrimination.
As an illustration of this approach, certain nonadaptive discrimination
has sometimes been discussed \cite{Jen-Pla-2016,Zhu-Pir-2020}.
However, an adaptive strategy would outperform the best nonadaptive strategy
except for some special cases
\cite{Chi-Dar-Per-2008,Hay-2009,Dua-Guo-Li-Li-2016,Pir-Lup-2017,Pir-Bar-Geh-Wee-2018,
Kat-Wil-2020,Puc-Paw-Kra-Kuk-2021}.
For example, it is known that there exist two channels that can be perfectly distinguished
by using an adaptive strategy with only two uses of the channel, while they
cannot be perfectly distinguished by using any nonadaptive strategy with a finite number of uses
\cite{Har-Has-Leu-Wat-2010}.
We present a lower bound that is obtained by an adaptive discrimination strategy
based on Bayesian updating.
Our work is motivated by the fact that, for quantum state discrimination,
a Bayesian updating approach has been shown to be effective
\cite{Bon-1993,Ass-Poz-Pie-2011,Bec-Fan-Bau-Gol-2013,Fla-Bar-Cro-2019}
and to be optimal at least for discriminating
two identical copies of a pure state \cite{Dol-1976,Bro-Mei-1996,Aci-Bag-Bai-Mas-2005}.
Our numerical results demonstrate the tightness of the proposed lower bound.
We should emphasize that the proposed upper and lower bounds
do not explicitly depend on any quantum phenomena, such as entanglement and quantum teleportation,
and can be readily extended to operational probabilistic theory (or generalized probabilistic theory)
\cite{Lud-1985,Har-Neu-1974,Bar-2007,Chi-Dar-Per-2010,Jan-Lal-2013}.

\begin{figure}[bt]
 \centering
 \InsertPDF{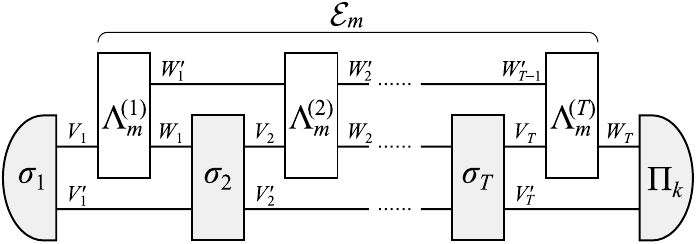}
 \caption{General protocol of quantum process discrimination.
 \RA{$\cE_m$ is a process consisting of $T$ channels $\Lambda^{(1)}_m,\dots,\Lambda^{(T)}_m$.
 Discrimination is characterized by the collection of a state $\sigma_1$,
 channels $\sigma_2,\dots,\sigma_T$, and a measurement $\{\Pi_k\}_{k=1}^M$.}}
 \label{fig:process_discrimination}
\end{figure}

\emph{Process discrimination problems} ---
Suppose that we want to discriminate between $M$ quantum processes $\cE_1,\dots,\cE_M$
as accurately as possible,
where each $\cE_m$ is a process consisting of $T$ channels $\Lambda^{(1)}_m,\dots,\Lambda^{(T)}_m$.
% \footnote{Our results can be easily extended to more general quantum operators
% (i.e., completely positive maps).}.
The most general discrimination protocol can be expressed as \RA{the collection
of a state $\sigma_1$, channels $\sigma_2,\dots,\sigma_T$,
and a measurement $\Pi \coloneqq \{\Pi_k\}_{k=1}^M$ (see Fig.~\ref{fig:process_discrimination}).}
Channels $\Lambda^{(t)}_m$ and $\Lambda^{(t+1)}_m$ are connected by an ancillary system $\W'_t$.
A process $\cE_m$, which is also called a quantum supermap or a quantum comb \cite{Chi-Dar-Per-2008},
is equivalent to a sequence of memory channels \cite{Chi-Dar-Per-2008-memory}.
In the first step of process discrimination,
a bipartite system $\V_1 \ot \V'_1$ is prepared in an initial state $\sigma_1$.
Its part $\V_1$ is sent through the channel $\Lambda^{(1)}_m$, followed by a channel $\sigma_2$.
Then, we send the system $\V_2$ through the channel $\Lambda^{(2)}_m$ and so on.
After $T$ steps, a quantum measurement $\Pi$ is performed
on the system $\W_T$.
The problem of discriminating $M$ channels $\Lambda_1,\dots,\Lambda_M$
with $T$ queries can be regarded as a special case of a processes discrimination problem
with $\Lambda^{(t)}_m = \Lambda_m$ and $\W'_t = \Complex$ $~(\forall m,t)$.
For simplicity, we focus on the case of equal prior probabilities.
\RA{Let $P_{k|m}$ be the conditional probability that the measurement outcome is $k$ given that
the given process is $\cE_m$, which is expressed by
\begin{alignat}{1}
 P_{k|m} \coloneqq \Pi_k \c \Lambda^{(T)}_m \c \sigma_T \c \cdots \c \sigma_2 \c \Lambda^{(1)}_m \c \sigma_1.
\end{alignat}
The success probability, $\PS$, is written as
\begin{alignat}{1}
 \PS &\coloneqq \frac{1}{M} \summ P_{m|m},
 \label{eq:PS}
\end{alignat}
}%
where $\c$ denotes function composition.
Our objective is to find discrimination $(\sigma_1,\dots,\sigma_T, \Pi)$ that
maximizes the success probability.
It is known that this optimization problem is formulated as a semidefinite programming (SDP)
problem of order $\tN \coloneqq \prod_{t=1}^T N_{\V_t} N_{\W_t}$ \cite{Chi-2012},
where \RA{$N_{\V_t}$ and $N_{\W_t}$ are, respectively, the dimensions of the systems $\V_t$ and $\W_t$}.
Solving this problem requires time polynomial in $\tN$,
and thus is generally intractable for large $T$.
Indeed, in the case of $N_{\V_t} = N_{\W_t} \eqqcolon N$ for each $t$, for example,
$\tN = N^{2T}$ is exponentially increasing with $T$.

\emph{Proposed upper bound} ---
The basic idea is quite simple: for each $t$, we only have to replace $\Lambda^{(t)}_m$
of Eq.~\eqref{eq:PS} by $s_t X_t$,
where $s_t$ and $X_t$ are, respectively, a positive real number and a channel
satisfying $s_t X_t \ge \Lambda^{(t)}_m$ $~(\forall m)$.
For two single-step processes $\Lambda$ and $\Lambda'$,
the inequality $\Lambda \ge \Lambda'$ denotes that
$\Lambda - \Lambda'$ is completely positive.
Such a pair $(s_t, X_t)$ obviously exists.
From Eq.~\eqref{eq:PS}, we have
\begin{alignat}{1}
 \PS &\le \frac{1}{M} \summ \Pi_m \c s_T X_T \c \sigma_T \c \cdots \c \sigma_2 \c s_1 X_1 \c \sigma_1
 = \frac{1}{M} \prod_{t=1}^T s_t,
\end{alignat}
where the equality follows from
$\summ \Pi_m \c X_T \c \sigma_T \c \cdots \c \sigma_2 \c X_1 \c \sigma_1 = 1$.
This gives that the ultimate success probability is upper bounded by $M^{-1} \prod_{t=1}^T s_t$.
For example, in the case of $T = 2$, it is diagrammatically depicted as
\begin{alignat}{1}
 \InsertPDF{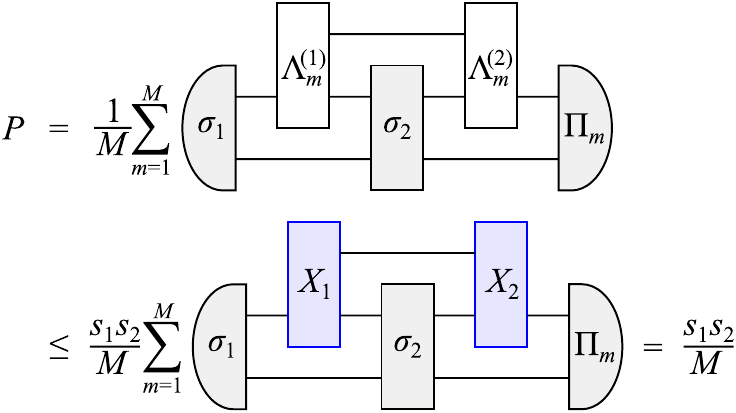} ~\raisebox{1em}{.}
\end{alignat}
To make this bound as tight as possible, we need to minimize $s_1,\dots,s_T$.
This problem is written as
\begin{alignat}{1}
 \begin{array}{ll}
  \mbox{minimize} & s_t \\
  \mbox{subject~to} & s_t X_t \ge \Lambda^{(t)}_m ~(\forall m) \\
 \end{array}
 \label{prob:s}
\end{alignat}
with a real number $s_t$ and a channel $X_t$ from $\W'_{t-1} \ot \V_t$ to $\W'_t \ot \W_t$.
Let $s_t^\opt$ be the optimal value of Problem~\eqref{prob:s}\RA{.
The proposed upper bound, $\oPS{1}$,} is \RA{given by}
\begin{alignat}{1}
 \oPS{1} &\coloneqq \frac{1}{M} \prod_{t=1}^T s_t^\opt\RA{,}
 \label{eq:bound_upper1}
\end{alignat}
\RA{where the subscript 1 indicates that $\oPS{1}$ is obtained by
optimization problems for finding single-step processes (the same for $\oPS{2}$,
which will be defined below).}

The above argument can be readily extended to obtain a tighter bound
at the expense of additional complexity.
For instance, instead of finding a single-step process $s_t X_t$ that
is larger than $\Lambda^{(t)}_m$ as in Eq.~\eqref{prob:s},
we can consider finding a pair of single-step processes that
is larger than the pair $[\Lambda^{(t-1)}_m,\Lambda^{(t)}_m]$.
Specifically, we consider the following optimization problem
\begin{alignat}{1}
 \begin{array}{ll}
  \mbox{minimize} & s_{t,2} \\
  \mbox{subject~to} & s_{t,2} X_t \c \eta \c X_{t-1} \ge \Lambda^{(t)}_m \c \eta \c \Lambda^{(t-1)}_m \\
  & ~(\forall m, \eta) \\
 \end{array}
 \label{prob:s2}
\end{alignat}
\RB{with a real number $s_{t,2}$ and channels $X_{t-1}$ and $X_t$,
which are the same type as $\Lambda^{(t-1)}_m$ and $\Lambda^{(t)}_m$, respectively.}
$\eta$ is any channel that can be sequentially connected to
channels $\Lambda^{(t)}_m$ and $\Lambda^{(t-1)}_m$ such as
$\Lambda^{(t)}_m \c \eta \c \Lambda^{(t-1)}_m$.
Its optimal solution, $s_{t,2}^\opt$, can be used to obtain an upper bound
instead of $s_{t-1}^\opt s_t^\opt$,
Thus, we obtain the following upper bound
\begin{alignat}{2}
 \oPS{2} &\coloneqq \frac{1}{M} \prod_{t=1}^{T/2} s_{2t,2}^\opt
 \quad \mbox{or} \quad
 \oPS{2} &\coloneqq \frac{s_T^\opt}{M} \prod_{t=1}^{(T-1)/2} s_{2t,2}^\opt,
 \label{eq:bound_upper2}
\end{alignat}
for even or odd $T$, respectively.
\RA{We can easily see $\PS \le \oPS{2} \le \oPS{1}$
\footnote{\RA{Let $(s_t^\opt,X_t^\opt)$ and $(s_{t-1}^\opt,X_{t-1}^\opt)$ be,
respectively, the optimal solutions to Eq.~\protect\eqref{prob:s}
and that with $t$ replaced by $t-1$; then,
we can easily verify that $(s_{t-1}^\opt s_t^\opt,X_{t-1}^\opt,X_t^\opt)$ is
a feasible solution to Eq.~\protect\eqref{prob:s2},
which yields $s_{t-1}^\opt s_t^\opt \ge s_{t,2}^\opt$.
Thus, $\oPS{2} \le \oPS{1}$ holds.
By the same discussion as $\PS \le \oPS{1}$, we obtain $\PS \le \oPS{2}$.}}.}

With the so-called Choi-Jamio{\l}kowski representation of $X_t$ \RA{\cite{Cho-1975,Jam-1972}},
Problems~\eqref{prob:s} and \eqref{prob:s2} can be formulated as SDP problems.
Thus, their numerical optimal solutions can be efficiently obtained
by several well-known SDP \RA{solvers}.
Analytical optimal solutions to these problems can be obtained in some cases,
such as the case in which processes $\cE_1,\dots,\cE_M$ have some kind of symmetry
\cite{Nak-Kat-2021-general};
another example is shown in \RO{\citesupF{supplemental:CPFAD}}.

As a special case, we consider the $T$-shot discrimination of quantum channels.
In this case, the optimal value $s_t^\opt \eqqcolon s^\opt$ of Problem~\eqref{prob:s}
is obviously independent of $t$.
Let $\PS^\opt_T$ be the ultimate success probability; then,
\RA{since $\PS^\opt_T$ increases as the number of evaluations $T$ increases,
$\PS^\opt_T \ge \PS^\opt_{T-1} \ge \cdots \ge \PS^\opt_1$ holds.}
As an application of the above argument, we also obtain
(see \RO{\citesup{supplemental:UB_channel}})
\begin{alignat}{1}
 \PS^\opt_T &\le s^\opt \PS^\opt_{T-1} \le s^\opt{}^2 \PS^\opt_{T-2} \le \cdots
 \le s^\opt{}^{T-1} \PS^\opt_1 = \frac{s^\opt{}^T}{M},
 \label{eq:PS_chain}
\end{alignat}
where the equality follows from $\PS^\opt_1 = s^\opt / M$ \cite{Chi-2012}.
This equation implies that
the ultimate success probability increases by at most $s^\opt \RA{~(\ge 1)}$ times
if the number of channel evaluations increases by one.
Discrimination of quantum channels that are very close to each other
is required in many application scenarios such as quantum illumination \cite{Llo-2008,Tan-Erk-Gio-Guh-2008}
and quantum reading \cite{Pir-2011}.
In such a case, since $s^\opt$ is very close to one,
the inequality $\PS^\opt_{T-1} \le \PS^\opt_T \le s^\opt \PS^\opt_{T-1}$
provides a strong constraint.
\RB{Equation~\eqref{eq:PS_chain} provides some useful properties.
As an example, we can see that the given channels cannot be perfectly discriminated
with $T$ uses if $\PS^\opt_1$ is smaller than $1/M^{1-1/T}$.
As another example, in order for the ultimate success probability
to be larger than a given threshold $p$,
more than $\log_{s^\opt} M p$ evaluations are needed.}

\emph{Proposed lower bound} ---
A natural approach for obtaining a lower bound is
to restrict attention to certain types of discrimination strategies.
A typical example is nonadaptive strategies.
The success probability, $P^\mathrm{(na)}$, of the best nonadaptive strategy
would be more easily obtained than the ultimate success probability;
for example, in the particular case of $T$-shot discrimination of
two channels $\Lambda_1$ and $\Lambda_2$,
it is well known that $P^\mathrm{(na)}$ is given by
$\frac{1}{2} + \frac{1}{4} \|\Lambda_1^{\ot T} - \Lambda_2^{\ot T}\|_\diamondsuit$.
However, adaptive strategies provide a clear advantage over nonadaptive ones in not a few cases.

\begin{figure}[bt]
 \centering
 \InsertPDF{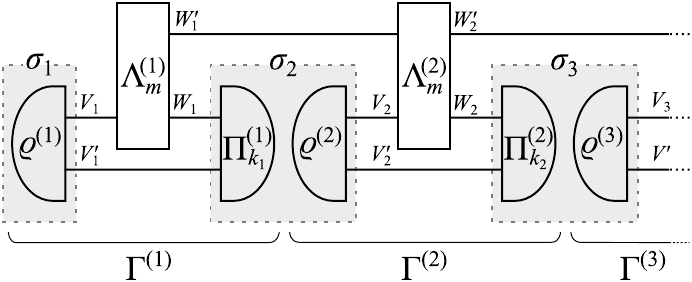}
 \caption{Proposed protocol based on Bayesian updating.
 Each channel $\sigma_t$ with $2 \le t \le T$ is
 restricted to a measure-and-prepare channel $\varrho^{(t)} \c \Pi^{(t-1)}$. 
 The state preparation $\varrho^{(t)}$ and the measurement $\Pi^{(t)}$ depend on the outcome of
 the previous measurement $\Pi^{(t-1)}$.
 To provide a tight lower bound on the ultimate success probability,
 $[\varrho^{(1)}, \Pi^{(1)}], [\varrho^{(2)}, \Pi^{(2)}], \dots$ are sequentially
 optimized.}
 \label{fig:bound_lower_Bayes}
\end{figure}

We propose an adaptive strategy based on Bayesian updating to obtain a tight lower bound.
In our method, channels $\sigma_2,\dots,\sigma_T$ are
restricted to measure-and-prepare (i.e., entanglement breaking) channels
as illustrated in Fig.~\ref{fig:bound_lower_Bayes}.
The channel $\sigma_t$ with $2 \le t \le T$ consists of
a measurement, $\Pi^{(t-1)} \coloneqq \{ \Pi^{(t-1)}_m \}_{m=1}^M$, followed by a
state preparation, $\varrho^{(t)}$.
\RB{The state preparation $\varrho^{(t)}$ and the measurement $\Pi^{(t)}$
can be connected by an ancillary system
and may depend on the outcome of the previous measurement $\Pi^{(t-1)}$.}
Assume that they are independent of the outcome of
measurements $\Pi^{(t-2)},\Pi^{(t-3)},\dots$ to reduce the complexity.
In such a scenario, we want to determine $\varrho^{(t)}$ and $\Pi^{(t)}$
such that the success probability is as high as possible.
For practical computation, we need to optimize them sequentially for $t = 1, 2, \dots$.
Note that, since such discrimination only requires state preparations and measurements,
it has the advantage of being relatively easy to implement experimentally.

We here present a brief outline of the proposed method;
we refer to \RO{\citesup{supplemental:Bayes}} for details.
\RB{Let $\varrho^{(1)} \coloneqq \sigma_1$ and $\Pi^{(T)} \coloneqq \Pi$; then,
the sequence of processes shown in Fig.~\ref{fig:bound_lower_Bayes}
is expressed by the sequential composition of
\begin{alignat}{1}
 \Gamma_{k_t|m,k_{t-1}}^{(t)} &\coloneqq \Pi^{(t)}_{k_t} \c \Lambda^{(t)}_m \c \varrho^{(t)},
\end{alignat}
where $k_{t-1}$ is the outcome of $\Pi^{(t-1)}$.
After some calculations, we find that
the probability that the measurement $\Pi^{(t)}$ correctly distinguishes
between the processes is given by
\begin{alignat}{1}
 P^{(t)} &\coloneqq \frac{1}{M} \summ q^{(t)}_m,
\end{alignat}
where $q^{(t)}_m$ is the conditional probability of
the outcome of the measurement $\Pi^{(t)}$ being $m$
given that the given process is $\cE_m$, which is expressed by
\begin{alignat}{1}
 q^{(t)}_m &\coloneqq \Tr \sum_{k_{t-1}=1}^M \dots \sum_{k_1=1}^M
 \Gamma_{m|m,k_{t-1}}^{(t)} \c \Gamma_{k_{t-1}|m,k_{t-2}}^{(t-1)}
 \c \cdots \c \Gamma_{k_1|m}^{(1)}.
\end{alignat}
The sets of $[\varrho^{(1)}, \Pi^{(1)}], [\varrho^{(2)}, \Pi^{(2)}], \dots$ can be
sequentially optimized.
Specifically, for each $t$, we find $[\varrho^{(t)}, \Pi^{(t)}]$ that maximize $P^{(t)}$,
which can be regarded as a single-shot channel discrimination problem
and is formulated as an SDP problem.}
The success probability of our strategy is given by
\begin{alignat}{1}
 \lPS &\coloneqq \RB{P^{(T)}},
 \label{eq:lPS}
\end{alignat}
which is obviously a lower bound on the ultimate success probability.
\RA{Since we need to optimize the $T$ sets
$[\varrho^{(1)}, \Pi^{(1)}], [\varrho^{(2)}, \Pi^{(2)}], \dots$,
the computational complexity of obtaining $\lPS$ is roughly proportional to $T$.}

We emphasize that since the proposed upper and lower bounds are based only on the concept of
an operational probabilistic framework,
it can be generalized to an arbitrary operational probabilistic theory.
In such a theory, we need to solve some convex programming problems that are not SDP in general.
However, these problems can be efficiently solved with existing techniques
such as interior-point methods.

\emph{Numerical results} ---
\begin{figure}[tb]
 \centering
 \includegraphics[scale=0.6]{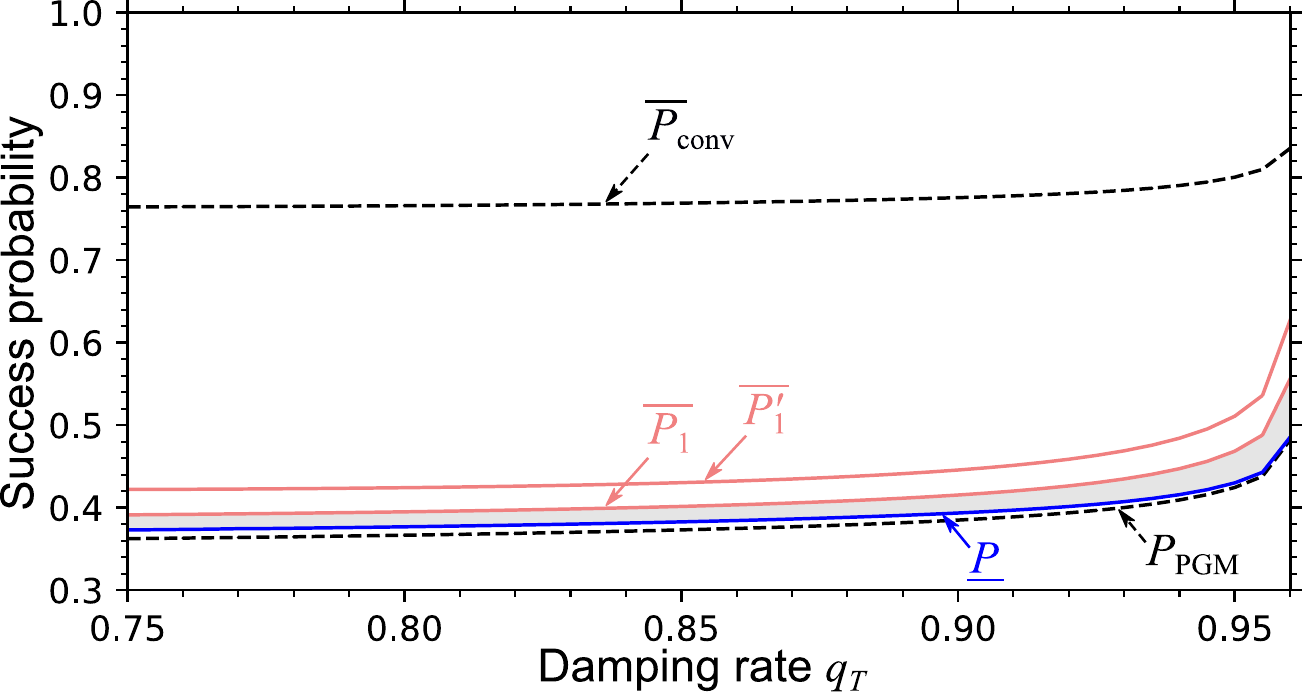}
 \caption{Success probability in the problem of channel position finding
 with two AD channels $\AD_{q_B}$ and $\AD_{q_T}$,
 where \RO{$T = 2$, $M = 3$,} and $q_B = q_T + 0.04$.
 The ultimate success probability lies in the gray region\RA{,
 between our upper bound $\oPS{1}$ of Eq.~\eqref{eq:bound_upper1} and
 our lower bound $\lPS$ of Eq.~\eqref{eq:lPS}.
 $\ol{P'_1}$ is the proposed upper bound described in the SM.
 $\ol{P}_\mathrm{conv}$ is the upper bound proposed in Ref.~\cite{Zhu-Pir-2020}.
 $P_\mathrm{PGM}$ is the success probability} achieved by the maximally entangled pure state
 and the pretty good measurement \cite{Hol-1978,Hau-Woo-1994},
 which is a lower bound on the ultimate success probability.}
 \label{fig:result-CPF}
 \includegraphics[scale=0.6]{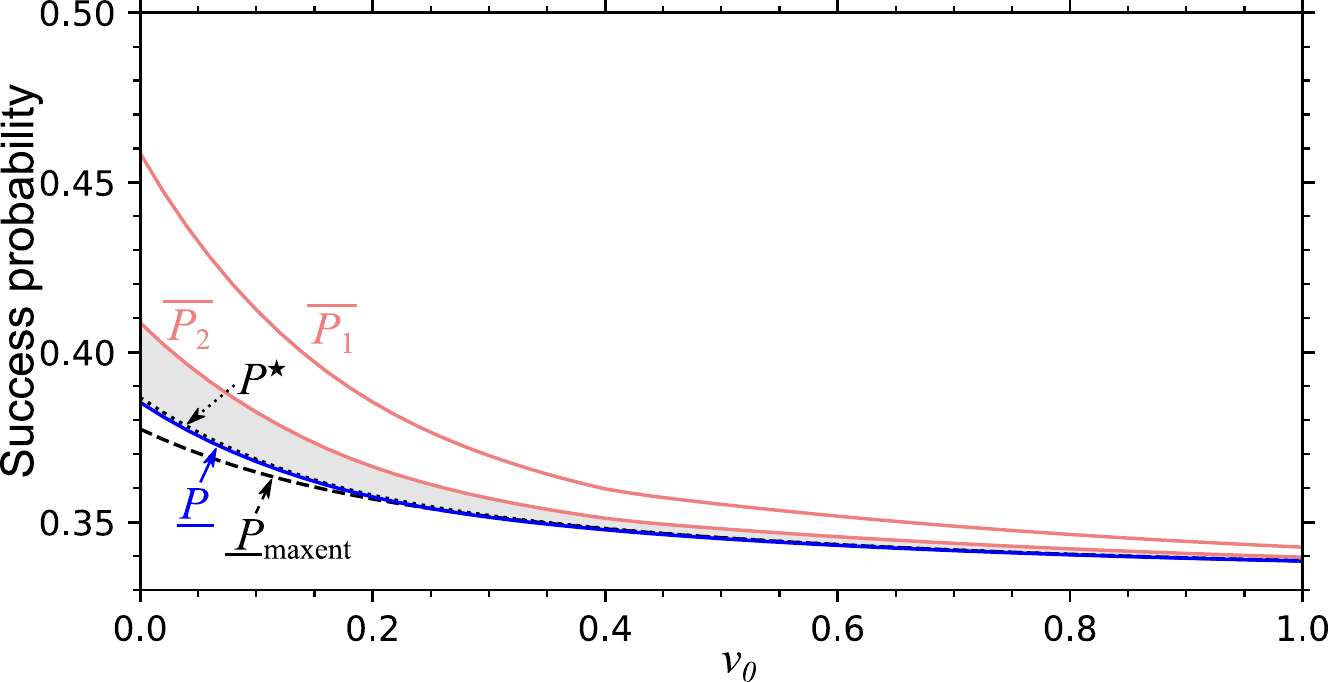}
 \caption{Success probability in the discrimination between
 processes consisting of two generalized AD channels with correlated noise,
 where $T = M = 3$.
 The parameters that are associated with the zero-temperature dissipation rate
 are set to $\nu_0$ and $\nu_0+0.04$, respectively.
 All the other parameters of these channels are the same.
 \RA{The two proposed upper bounds $\oPS{1}$ of Eq.~\eqref{eq:bound_upper1} and
 $\oPS{2}$ of Eq.~\eqref{eq:bound_upper2} and
 the proposed lower bound $\lPS$ of Eq.~\eqref{eq:lPS} are depicted.}
 In this case, we can numerically compute the ultimate success probability $\PS^\opt$.
 \RA{$\lPS_\mathrm{maxent}$ is the success probability}
 achieved by the maximally entangled pure state and
 the optimal measurement, which gives a lower bound on $\PS^\opt$.}
 \label{fig:result-GAD_memory}
\end{figure}
First, we discuss a multi-shot channel discrimination problem.
We here consider the problem of channel position finding \cite{Zhu-Pir-2020-entangle}
with two amplitude damping (AD) channels to compare our results with
that in Ref.~\cite{Zhu-Pir-2020}.
\RA{Let $\AD_q$ be} the AD channel with the damping parameter $q$,
i.e., the qubit channel defined by
\begin{alignat}{1}
 \AD_q(\rho) &= E_0 \rho E_0^\dagger + E_1 \rho E_1^\dagger, \nonumber \\
 E_0 &\coloneqq \ket{0}\bra{0} + \sqrt{1-q} \ket{1}\bra{1}, \quad
 E_1 \coloneqq \sqrt{q}\ket{0}\bra{1}
 \label{eq:AD}
\end{alignat}
with the standard basis $\{ \ket{0}, \ket{1} \}$.
Specifically, we consider $T$-shot discrimination of three channels,
in which case the three channels are \RA{expressed in the form}
$\AD_{q_T} \otimes \AD_{q_B} \otimes \AD_{q_B}$,
$\AD_{q_B} \otimes \AD_{q_T} \otimes \AD_{q_B}$,
and $\AD_{q_B} \otimes \AD_{q_B} \otimes \AD_{q_T}$
\RA{with two damping parameters $q_T$ and $q_B$}.
In Fig.~\ref{fig:result-CPF}, we show our numerical results.
\RA{We computed our bounds $\oPS{1}$ and $\lPS$ from Eqs.~\eqref{eq:bound_upper1}
and \eqref{eq:lPS}, respectively,
where we solved the corresponding single-shot channel discrimination problems
by the SDP solver CSDP \cite{Bor-1999}.
We also computed another proposed upper bound $\ol{P'_1}$
(detailed in \citesup{supplemental:UB_ext}),
which can be obtained at low computational cost.
We can see that $\ol{P}_\mathrm{conv}$ is far from being optimal
when the given channels are very close to each other.
Indeed, $\ol{P}_\mathrm{conv} \ge (M+1)/2M$ always holds
for any discrimination problem of $M$ channels with equal prior probabilities
\footnote{\RA{From Eq.~(7) of Ref.~\cite{Zhu-Pir-2020}, we have
$\ol{P}_\mathrm{conv} \ge 1 - (M-1)/2M = (M+1)/2M$.}},
while the ultimate success probability $P^\opt$ is close to $1/M$
when the given channels are nearly identical to each other.
We can say that $\oPS{1}$ is tighter than $\ol{P}_\mathrm{conv}$ in such a situation
if $T$ is not large enough
\footnote{\RA{Let $\PS^\opt_1$ be the ultimate success probability in the case of $T = 1$;
then, $\oPS{1} = M^{T-1} \PS^\opt_1{}^T$ holds from
$\oPS{1} = s^\opt{}^T/M$ and $\PS^\opt_1 = s^\opt/M$,
where $s^\opt$ is the optimal value of Problem~\protect\eqref{prob:s}.
It follows from $\ol{P}_\mathrm{conv} \ge (M+1)/2M$ that
$\oPS{1}$ is tighter than $\ol{P}_\mathrm{conv}$
whenever $\PS^\opt_1 < [(M+1)/2]^{1/T}/M$.}}.
Note that the proposed bound $\oPS{1}$ becomes looser as $T$ increases;
in our preliminary numerical experiments, we observed that $\oPS{1}$ is worse than
$\ol{P}_\mathrm{conv}$ for large $T$ (e.g., $T \ge 13$).}
As for the computational cost, computing $\oPS{1}$ requires $\RO{\poly(4^M)}$ time,
whereas $\ol{P}_\mathrm{conv}$ requires $\RO{O(1)}$ time
\RA{(see \citesup{supplemental:complexity})}.
The proposed method can be easily extended to obtain
a slightly looser bound $\ol{P'_1}$ requiring $\RO{O(1)}$ time.
Computing $P_\mathrm{PGM}$ and $\lPS$ takes $\RO{\poly(4^{MT})}$ and
$\RO{O(MT) \poly(4^M)}$ times, respectively
\footnote{\RA{We did this numerical experiment on a PC with 16~GB memory,
in which case neither $\oPS{1}$ for $M \ge 4$ nor $P_\mathrm{PGM}$
for $TM \ge 9$ can be computed due to memory limitations.}}.

Next, we discuss the problem of discriminating $M$ processes
\RB{where each process $\cE_m$ consists of $M$ memory channels
each of which is the same channel, $\GAD_0$, except the $m$-th step, which is $\GAD_1$.}
These processes are analogous to pulse-position modulated signals;
We are here concerned with the case in which \RB{$\GAD_0$ and $\GAD_1$ are memory channels
each of which} is associated with two consecutive uses of generalized AD channel
with correlated noise.
Additional details including the exact definition of generalized AD channels
are given in \RO{\citesup{supplemental:numericalGAD}}.
Figure~\ref{fig:result-GAD_memory} shows the two upper bounds $\oPS{1}$ of
Eq.~\eqref{eq:bound_upper1} and $\oPS{2}$ of Eq.~\eqref{eq:bound_upper2} and
the lower bound $\lPS$.
\RA{In this simulation, we set $M = 3$ to compute the exact value of
the ultimate success probability $\PS^\opt$
(note that $T = M$ holds in this problem).
Since the cost of computing $\PS^\opt$
increases exponentially with $M$, $\PS^\opt$ is practically computable
only for fairly small $M$ (typically, $M \le 3$).}
\RA{We observe that $\lPS$ is very close to $\PS^\opt$;
the difference between them is less than 0.0015.}
$\oPS{1}$, $\oPS{2}$, and $\lPS$ have affordable computational costs;
they require $\RO{O(1)}$, $\RO{O(1)}$, and $\RO{O(M^2)}$ times, respectively.

\emph{Conclusions} ---
We presented upper and lower bounds on the ultimate success probability
for discriminating arbitrary finite-dimensional quantum processes.
In a special case of multi-shot channel discrimination,
the ultimate success probability satisfies 
the relationship of Eq.~\eqref{eq:PS_chain}.
Our approach can be used to estimate the ultimate performances
in various quantum information tasks,
such as quantum sensing, quantum imaging, and quantum tomography.

We thank for O.~Hirota and T.~S.~Usuda for comments and discussions.
This work was supported by JSPS KAKENHI Grant Number JP19K03658.

\bibliographystyle{apsrev4-1}

% \bibliography{quant}

%%%%%%%%%%%%%%%%%%%%%%%%%%%%%%%%%%%%%%%%%%%%%%%%%%%%%%%%%%%%%%%%%%%%%%%%%%%%%%%%
% Supplemental Material

\clearpage\onecolumngrid

\pagestyle{empty}
\setcounter{equation}{0}
\renewcommand{\theequation}{S\arabic{equation}}
\setcounter{table}{0}
\renewcommand{\thetable}{S\arabic{table}}
\setcounter{section}{0}
\renewcommand{\citenumfont}[1]{S#1}
\renewcommand\bibnumfmt[1]{[S#1]}

\begin{center}
 \supplementaltitle%
 \supplementalauthor{Kenji Nakahira and Kentaro Kato}%
 \supplementalaffiliation{%
 Quantum Information Science Research Center, Quantum ICT Research Institute, Tamagawa University,
 Machida, Tokyo 194-8610, Japan
 }%
\end{center}

\section{Notation} \label{supplemental:notation}

We first introduce some notation.
Let $\Real_+$ and $\Complex$ be, respectively, the sets of all
nonnegative real numbers and all complex numbers.
We will identify a one-dimensional system with $\Complex$.
We denote by $\NV$ and $\{ \ket{n} \}_{n=0}^{\NV-1}$, respectively,
the dimension and the standard basis of $\V$.
$\Pos(\V,\W)$ and $\Chn(\V,\W)$, respectively, denote the sets of all single-step processes
(i.e., completely positive maps) and
channels (i.e., trace-preserving completely positive maps) from a system $\V$ to a system $\W$.
Let $\ident_\V \in \Chn(\V,\V)$ be the identity map on a system $\V$.
Let $\Pos_\V$ be the set of all positive semidefinite matrices on $\V$
and $\Den_\V$ be the set of all elements of $\Pos_\V$ with unit trace
(i.e., density matrices on $\V$).
$\I_\V \in \Pos_\V$ denotes the identity matrix on $\V$.
A process $\cE$ that consists of $T \ge 1$ time steps is expressed in the form
\begin{alignat}{1}
 \cE &= \Lambda^{(T)} \ast \cdots \ast \Lambda^{(1)},
 \quad \Lambda^{(t)} \in \Chn(\W'_{t-1} \ot \V_t, \W'_t \ot \W_t),
 \quad \W'_0 = \W'_T = \Complex,
 \label{eq:cE_Lambda0}
\end{alignat}
where $\ast$ is the link product \cite{Chi-Dar-Per-2008-sm}.
$\cE$ is diagrammatically depicted as
(see Fig.~\ref{fig:process_discrimination} in the main paper)
\begin{alignat}{1}
 \InsertPDF{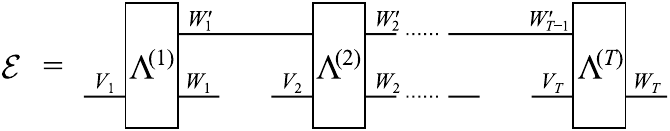} ~\raisebox{.5em}{.}
\end{alignat}
Let
\begin{alignat}{1}
 \InsertPDF{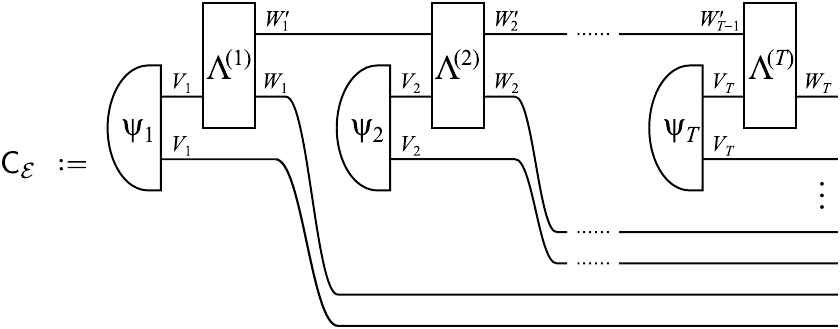} ~\raisebox{.0em}{,}
\end{alignat}
where $\Psi_t \coloneqq \ket{\Psi_t}\bra{\Psi_t}$ and
$\ket{\Psi_t} \coloneqq \sum_{n=0}^{N_{\V_t}-1} \ket{n} \ot \ket{n} \in \V_t \ot \V_t$.
We can easily verify $\C_\cE \in \Pos_{\W_T \ot \V_T \ot \cdots \W_1 \ot \V_1}$.
In the particular case of $\cE \in \Pos(\V,\W)$,
it is easily seen that $\C_\cE \in \Pos_{\W \ot \V}$ is its unnormalized Choi matrix.
\RO{Equation and figure numbers without the prefix `S' refer to those given in the main paper.}

\section{Proposed upper bound} \label{supplemental:UB}

\subsection{Formulation} \label{supplemental:UB_formulation}

We will be concerned with the problem of discriminating $M$ processes $\cE_1,\dots,\cE_M$
that consist of $T$ time steps.
Assume that, for each $m$, $\cE_m$ is expressed in the form
\begin{alignat}{1}
 \cE_m &\coloneqq \Lambda^{(T)}_m \ast \cdots \ast \Lambda^{(1)}_m,
 \quad \Lambda^{(t)}_m \in \Chn(\W'_{t-1} \ot \V_t, \W'_t \ot \W_t),
 \quad \W'_0 = \W'_T = \Complex.
 \label{eq:cE_Lambda}
\end{alignat}
Let $p_1,\dots,p_M$ be their prior probabilities.
Let us choose $h_1,\dots,h_L$ such that
\begin{alignat}{1}
 0 = h_0 < h_1 < h_2 < \dots < h_L = T
\end{alignat}
and rewrite $\cE_m$ as
\begin{alignat}{1}
 \cE_m &= \cE^{(L)}_m \ast \cE^{(L-1)}_m \ast \cdots \ast \cE^{(1)}_m, \nonumber \\
 \cE^{(l)}_m &\coloneqq \Lambda^{(h_l)}_m \ast \Lambda^{(h_l-1)}_m \ast \cdots \ast \Lambda^{(h_{l-1}+1)}_m,
 \quad \forall l \in \{1,\dots,L\}.
 \label{eq:cE_L}
\end{alignat}
It is easily seen that $\cE^{(l)}_m$ is a $(h_l-h_{l-1})$-step process.
Arbitrarily choose $p^{(l)}_m \in \Real_+$ satisfying
$\prod_{l=1}^L p^{(l)}_m = p_m$.
For each $l \in \{1,\dots,L\}$, let us consider the following optimization problem:
\begin{alignat}{1}
 \begin{array}{ll}
  \mbox{minimize} & s_l \\
  \mbox{subject~to} &
   s_l X_{h_l} \c \eta_{h_l} \c \cdots \c X_{h_{l-1}+2} \c \eta_{h_{l-1}+2} \c X_{h_{l-1}+1}
   \ge p^{(l)}_m \Lambda^{(h_l)}_m \c \eta_{h_l} \c \cdots \c \Lambda^{(h_{l-1}+2)}_m
   \c \eta_{h_{l-1}+2} \c \Lambda^{(h_{l-1}+1)}_m \\
  & [\forall m \in \{1,\dots,M\},
   \{ \eta_t \in \Chn(\W_{t-1} \ot \V'_{t-1}, \V_t \ot \V'_t) \}_{t=h_{l-1}+2}^{h_l}] \\
 \end{array}
 \label{prob:sL}
\end{alignat}
with $s_l \in \Real_+$
and $\{ X_t \in \Chn(\W'_{t-1} \ot \V_t, \W'_t \ot \W_t) \}_{t=h_{l-1}+1}^{h_l}$,
where $\V'_t$ is any system.
Note that the constraint is equivalent to
\begin{alignat}{1}
 s_l \C_{X_{h_l} \ast \cdots \ast X_{h_{l-1}+1}}
 \ge p^{(l)}_m \C_{\cE^{(l)}_m}, \quad \forall m \in \{1,\dots,M\}.
\end{alignat}
Let $\PS$ be the success probability of discrimination $(\sigma_1,\dots,\sigma_T, \Pi)$.
We have that, for any feasible solutions $s_1,\dots,s_L$ and $X_1,\dots,X_T$
to Problem~\eqref{prob:sL},
\begin{alignat}{1}
 \PS &= \summ p_m \Pi_m \c \Lambda^{(T)}_m \c \sigma_T \c \cdots
 \c \sigma_2 \c \Lambda^{(1)}_m \c \sigma_1 \nonumber \\
 &= \summ \Pi_m \c \left[ p^{(L)}_m \Lambda^{(h_L)}_m \c \sigma_{h_L} \c \cdots
 \c \Lambda^{(h_{L-1}+1)}_m \right] \c \sigma_{h_{L-1}+1} \c
 \left[ p^{(L-1)}_m \Lambda^{(h_{L-1})}_m \c \sigma_{h_{L-1}} \c \cdots
 \c \Lambda^{(h_{L-2}+1)}_m \right] \nonumber \\
 &\quad \c \cdots \c \sigma_{h_1+1} \c
 \left[ p^{(1)}_m \Lambda^{(h_1)}_m \c \sigma_{h_1} \c \cdots
 \c \Lambda^{(1)}_m \right] \c \sigma_1 \nonumber \\
 &\le \summ \Pi_m \c \left[ s_L X_{h_L} \c \sigma_{h_L} \c \cdots \c X_{h_{L-1}+1} \right] \c
 \sigma_{h_{L-1}+1} \c \left[ s_{L-1} X_{h_{L-1}} \c \sigma_{h_{L-1}} \c \cdots
 \c X_{h_{L-2}+1} \right] \nonumber \\
 &\quad \c \cdots \c \sigma_{h_1+1} \c
 \left[ s_1 X_{h_1} \c \sigma_{h_1} \c \cdots \c X_1 \right] \c \sigma_1 \nonumber \\
 &= \prod_{l=1}^L s_l \summ \Pi_m \c X_T \c \sigma_T \c \cdots
 \c \sigma_2 \c X_1 \c \sigma_1 \nonumber \\
 &= \prod_{l=1}^L s_l,
\end{alignat}
where the last line follows since $\summ \Pi_m \c X_T \c \sigma_T \c \cdots
\c \sigma_2 \c X_1 \c \sigma_1 = 1$ holds for any
discrimination $(\sigma_1,\dots,\sigma_T, \Pi)$.
Thus, $\PS$ is upper bounded by $\prod_{l=1}^L s_l$.
To obtain the tightest bound, we may choose $s_l$ as the optimal solution to Problem~\eqref{prob:sL}.
Note that using a feasible solution instead of the optimal one would be practical
if Problem~\eqref{prob:sL} is too hard to solve.
In the special case of $L = 1$, the optimal value of $s_1$ is equal to
the ultimate success probability \cite{Chi-2012-sm}.
The upper bound $\oPS{1}$ of Eq.~\eqref{eq:bound_upper1} can be regarded
as the special cases of $L = T$ and $h_l = l$.
Similarly, the upper bound $\oPS{2}$ of Eq.~\eqref{eq:bound_upper2} can be regarded as
the case of $L = T/2$ and $h_l = 2l$ (for even $T$)
or $L = (T+1)/2$, $h_l = 2l$ $~(l=1,\dots,L-1)$, and $h_L = T$ (for odd $T$).

By substituting
\begin{alignat}{1}
 \chi_{l,t} &\coloneqq s_l \C_{X_t \ast \cdots \ast X_{h_{l-1}+1}}
 \in \Pos_{\W_t \ot \V_t \ot \W_{t-1} \ot \V_{t-1} \ot
 \cdots \ot \W_{h_{l-1}+1} \ot \V_{h_{l-1}+1}},
 \quad t \in \{ h_{l-1},\dots,h_l \}
\end{alignat}
into Problem~\eqref{prob:sL}, we can rewrite this problem as the following
semidefinite programming (SDP) problem \cite{Chi-2012-sm}:
\begin{alignat}{1}
 \begin{array}{ll}
  \mbox{minimize} & \chi_{l,h_{l-1}} \in \Real_+ \\
  \mbox{subject~to} & \Trp{\W_t} \chi_{l,t} = I_{\V_t} \ot \chi_{l,t-1}
   ~(\forall t \in \{h_{l-1}+1,\dots,h_l\}) \\
  & \chi_{l,h_l} \ge p^{(l)}_m \C_{\cE^{(l)}_m} ~(\forall m \in \{1,\dots,M\}) \\
 \end{array}
 \label{prob:sL_chi}
\end{alignat}
with $\{ \chi_{l,t} \}_{t=h_{l-1}}^{h_l}$.
Since $\chi_{l,h_l}$ is a positive semidefinite matrix of order
$\tN^{(l)} \coloneqq \prod_{t=h_{l-1}+1}^{h_l} N_{\V_t} N_{\W_t}$,
solving this problem requires time polynomial in $\tN^{(l)}$.

\subsection{Multi-shot channel discrimination} \label{supplemental:UB_channel}

As a special case of process discrimination,
let us consider $T$-shot discrimination of quantum channels $\{ \Lambda_m \}_{m=1}^M$
with prior probabilities $\{ p_m \}_{m=1}^M$.
This can be interpreted as the case of $\Lambda^{(1)}_m = \cdots = \Lambda^{(T)}_m = \Lambda_m$.
The ultimate success probability will be denoted by $\PS^\opt_T$, which is a function of $T$.
Let $\PS^\bullet_T$ be the ultimate success probability in the case of equal prior probabilities.
We will denote by $s_l^\opt$ the optimal value of Problem~\eqref{prob:sL}.
Let us substitute $p^{(l)}_m = 1$ $~(\forall l < L)$ and $p^{(L)}_m \coloneqq p_m$
into Problem~\eqref{prob:sL};
then, for each $l < L$, $s_l^\opt$ is equal to $M$ times
the ultimate success probability of $(h_l-h_{l-1})$-shot discrimination of
channels $\{ \Lambda_m \}$ with equal prior probabilities,
which gives $s_l^\opt = M \PS^\bullet_{h_l-h_{l-1}}$.
Also, $s_L^\opt$ is equal to the ultimate success probability of
$(h_L-h_{L-1})$-shot discrimination of channels $\{ \Lambda_m \}$ with the prior
probabilities of $\{ p_m \}$, i.e., $s_L^\opt = \PS^\opt_{h_L-h_{L-1}}$ holds.
Thus, we have
\begin{alignat}{2}
 \PS^\opt_T &\le \prod_{l=1}^L s_l^\opt &&= \left( \prod_{l=1}^{L-1} M \PS^\bullet_{h_l-h_{l-1}} \right)
 \PS^\opt_{h_L-h_{L-1}}.
\end{alignat}
In the special case of $L = 2$ and $h_1 = 1$,
we obtain $\PS^\opt_T \le (M \PS^\bullet_1) \PS^\opt_{T-1}$.
Applying this inequality recursively gives
\begin{alignat}{2}
 \PS^\opt_T &\le (M \PS^\bullet_1) \PS^\opt_{T-1} &&\le (M \PS^\bullet_1)^2 \PS^\opt_{T-2}
 \le \dots \le (M \PS^\bullet_1)^{T-1} \PS^\opt_1.
\end{alignat}
Equation~\eqref{eq:PS_chain} is a special case of this inequality.

\subsection{Extension of the proposed upper bound} \label{supplemental:UB_ext}

Our method presented in Subsec.~\ref{supplemental:UB_formulation}
can be intuitively understood as an approach based on the formulation of
each candidate process $\cE_m$ as the sequential composition of some small processes
$\cE^{(1)}_m,\dots,\cE^{(L)}_m$.
This is easily extended to the case in which each $\cE^{(l)}_m$ can be
expressed as the parallel composition (i.e., the tensor product) of some partitions.

Let us consider a process $\cE_m$ expressed by Eq.~\eqref{eq:cE_Lambda}.
Assume that, for each $m \in \{1,\dots,M\}$ and $t \in \{1,\dots,T\}$,
$\Lambda^{(t)}_m$ can be expressed in the form
\begin{alignat}{1}
 \Lambda^{(t)}_m &= \Lambda^{(t,1)}_m \ot \Lambda^{(t,2)}_m \ot \cdots \ot \Lambda^{(t,J_t)}_m,
\end{alignat}
where $\Lambda^{(t,j)}_1,\dots,\Lambda^{(t,j)}_M$ are of the same type
(i.e., they have the same input and output systems).
For simplicity, we will consider only the case of $L = T$ and $h_l = l$
in Eq.~\eqref{eq:cE_L} (in which case $\cE^{(t)}_m = \Lambda^{(t)}_m$ holds),
but can be easily extended to more general cases.
Let us arbitrarily choose a nonnegative real number $p^{(t,j)}_m$
such that $\prod_{t=1}^T \prod_{j=1}^{J_t} p^{(t,j)}_m = p_m$.
Instead of Problem~\eqref{prob:sL}, we consider, for each $t$ and $j$,
the following problem:
\begin{alignat}{1}
 \begin{array}{ll}
  \mbox{minimize} & s_{t,j} \\
  \mbox{subject~to} & s_{t,j} X_{t,j} \ge p^{(t,j)}_m \Lambda^{(t,j)}_m ~(\forall m) \\
 \end{array}
 \label{prob:sj}
\end{alignat}
with $s_{t,j} \in \Real_+$ and $X_{t,j}$,
where $X_{t,j}$ is a channel with the same type as $\Lambda^{(t,j)}_m$.
For example, in the case of $T = 3$ and $(J_1,J_2,J_3) = (3,1,2)$,
it follows that any feasible solution $(s_{t,j}, X_{t,j})$ to Problem~\eqref{prob:sj} and
any discrimination $(\sigma_1,\sigma_2,\sigma_3,\Pi)$ satisfy
\begin{alignat}{1}
 \InsertPDF{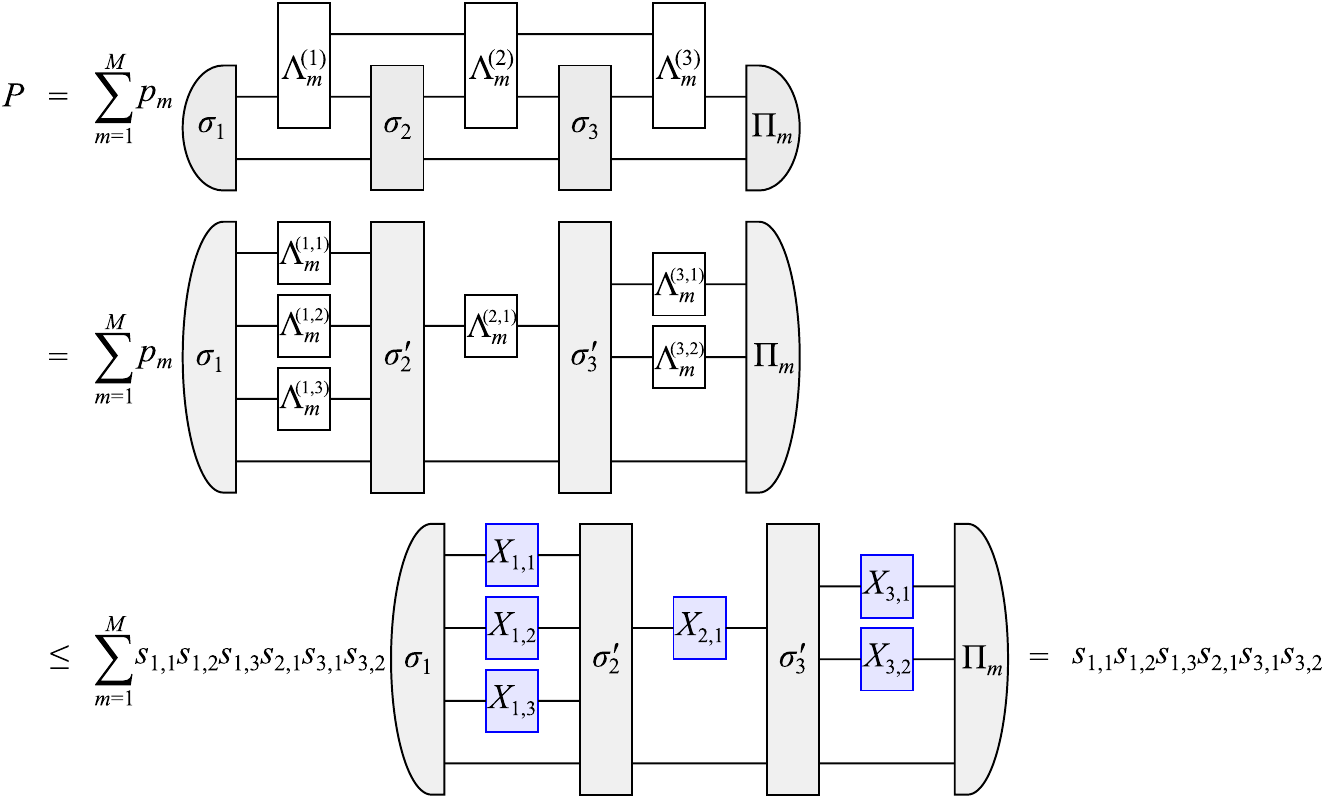} ~\raisebox{3.5em}{,}
\end{alignat}
where $\sigma'_t \coloneqq \ident_{\W'_{t-1}} \ot \sigma_t$.
This implies that the success probability is upper bounded by
$\prod_{t=1}^T \prod_{j=1}^{J_t} s_{t,j}$.
When choosing $s_{t,j}$ as the optimal value, $s_{t,j}^\opt$,
of Problem~\eqref{prob:sj}, we obtain the tightest bound
$\ol{P'_1} \coloneqq \prod_{t=1}^T \prod_{j=1}^{J_t} s_{t,j}^\opt$.
Note that $\ol{P'_1}$ is not tighter than $\oPS{1} \coloneqq \prod_{l=1}^T s_l^\opt$,
where $s_l^\opt$ is the optimal value of Problem~\eqref{prob:sL}
with $L = T$, $h_l = l$, and $p^{(t)}_m = \prod_{t=1}^{J_t} p^{(t,j)}_m$.
Indeed, this follows from
$\left( \prod_{j=1}^{J_t} s_{t,j}, \bigotimes_{j=1}^{J_t} X_{t,j} \right)$
being a feasible solution to Problem~\eqref{prob:sL}.
We often obtain $\ol{P'_1}$ with a much lower computational cost than $\oPS{1}$.

As an example, let us consider the problem of channel position finding
with amplitude damping (AD) channels \cite{Zhu-Pir-2020-sm},
i.e., the problem of $T$-shot discrimination of $M$ channels $\{ \tLambda_m \}_{m=1}^M$
with equal prior probabilities,
where $\tLambda_m$ is expressed as
\begin{alignat}{1}
 \tLambda_m &\coloneqq \Lambda_{\delta_{m,1}} \ot \Lambda_{\delta_{m,2}} \ot \cdots \ot
 \Lambda_{\delta_{m,M}}.
 \label{eq:CPF_Lambda}
\end{alignat}
$\Lambda_0 \coloneqq \AD_{q_B}, \Lambda_1 \coloneqq \AD_{q_T} \in \Chn(\V,\W)$
are AD channels, which is defined by Eq.~\eqref{eq:AD},
where $\V$ and $\W$ are qubit systems.
For example, we have $\tLambda_2 = \AD_{q_B} \ot \AD_{q_T} \ot \AD_{q_B}
\ot \cdots \ot \AD_{q_B}$.
Let us consider the following optimization problem:
\begin{alignat}{1}
 \begin{array}{ll}
  \mbox{minimize} & s \\
  \mbox{subject~to} & s X \ge \AD_{q_B}, ~s X \ge \AD_{q_T}
 \end{array}
 \label{prob:sAD}
\end{alignat}
with $s \in \Real_+$ and $X \in \Chn(\V,\W)$.
As an example, in the case of $M = T = 3$, it is easily seen that
the success probability $\PS$ of discrimination $(\sigma_1,\sigma_2,\sigma_3,\Pi)$
satisfies
\begin{alignat}{1}
 \InsertPDF{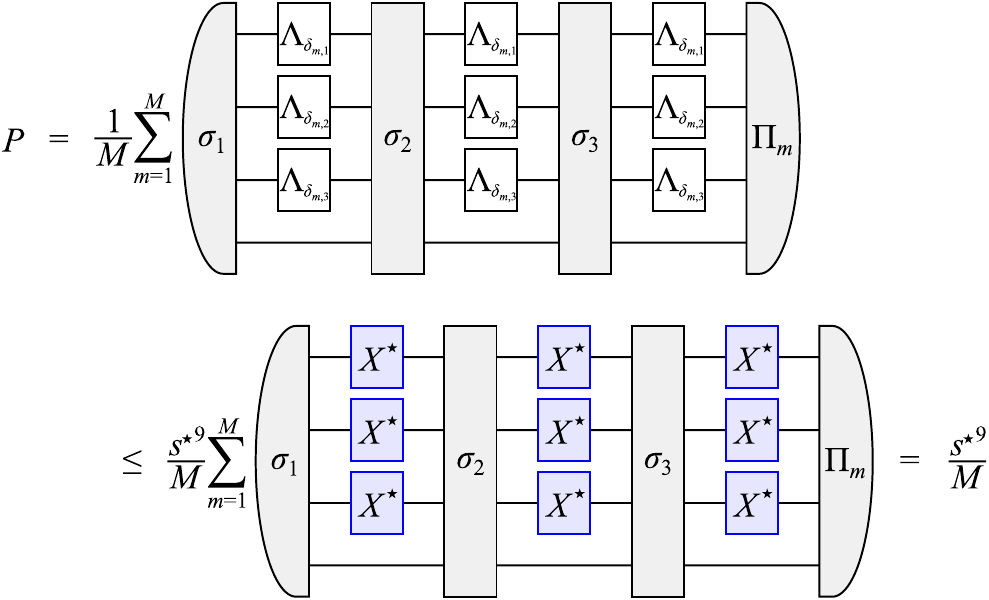} ~\raisebox{3.2em}{,}
\end{alignat}
where $(s^\opt, X^\opt)$ is the optimal solution to Problem~\eqref{prob:sAD}.
Thus, $\ol{P'_1} \coloneqq s^{\opt 9} / M$ is an upper bound on the ultimate success probability.
$\ol{P'_1}$ is often reasonably tight
(e.g., see Fig.~\ref{fig:result-CPF}).

\section{Discrimination of binary amplitude damping channels} \label{supplemental:CPFAD}

We show that Problem~\eqref{prob:sAD}, which is an example of Problem~\eqref{prob:s},
can be solved analytically.
Without loss of generality, assume $q_B > q_T$.
Let $Y \coloneqq s \C_X \in \Pos_{\W \ot \V}$; then,
Problem~\eqref{prob:sAD} is rewritten as
\begin{alignat}{1}
 \begin{array}{ll}
  \mbox{minimize} & s \\
  \mbox{subject~to} & \Trp{\W} Y = s \I_\V, ~Y \ge \C_{\AD_{q_B}}, ~Y \ge \C_{\AD_{q_T}}.
 \end{array}
 \label{prob:sAD2}
\end{alignat}
Let us denote by $(s^\opt, Y^\opt)$ its optimal solution.
$\C_{\AD_q}$ is expressed as
\begin{alignat}{1}
 \C_{\AD_q} &=
 \begin{bmatrix}
  1 & 0 & 0 & \sqrt{1-q} \\
  0 & q & 0 & 0 \\
  0 & 0 & 0 & 0 \\
  \sqrt{1-q} & 0 & 0 & 1-q \\
 \end{bmatrix}.
\end{alignat}
Thus, it follows that $Y^\opt$ is expressed in the form
\begin{alignat}{1}
 Y^\opt &=
 \begin{bmatrix}
  s^\opt & 0 & 0 & y_2 \\
  0 & y_1 & 0 & 0 \\
  0 & 0 & 0 & 0 \\
  y_2 & 0 & 0 & s^\opt - y_1 \\
 \end{bmatrix},
\end{alignat}
where $y_1$ and $y_2$ are real numbers.
From $q_B > q_T$, we can assume $y_1 = q_B$.
Thus, we only have to find the minimum $s^\opt$ and $y_2$ such that
\begin{alignat}{1}
 \begin{bmatrix}
  s^\opt & y_2 \\
  y_2 & s^\opt - q_B \\
 \end{bmatrix}
 &\ge
 \begin{bmatrix}
  1 & \sqrt{1-q_B} \\
  \sqrt{1-q_B} & 1-q_B \\
 \end{bmatrix}
 , \quad
 \begin{bmatrix}
  s^\opt & y_2 \\
  y_2 & s^\opt - q_B \\
 \end{bmatrix}
 \ge
 \begin{bmatrix}
  1 & \sqrt{1-q_T} \\
  \sqrt{1-q_T} & 1-q_T \\
 \end{bmatrix}.
\end{alignat}
After performing some algebra, we obtain
\begin{alignat}{1}
 s^\opt &=
 \begin{dcases}
  1 + q_B - q_T, & q_T < q_\mathrm{th}, \\
  1 - \sqrt{1-q_B}
  + \frac{(q_B - q_T) \left( 1 - \sqrt{1-q_B} \right)}{2\sqrt{1-q_T} + 2\sqrt{1-q_B} - q_B + q_T},
  & q_T \ge q_\mathrm{th}, \\
 \end{dcases} \nonumber \\
 q_\mathrm{th} &\coloneqq 1 - \frac{(1+q_B-q_T)^2}{4}.
\end{alignat}

\section{Proposed lower bound} \label{supplemental:Bayes}

As a preliminary, we consider the problem of single-shot discrimination of
channels $\{ \Lambda_m \in \Chn(\V,\W) \}_{m=1}^M$
with prior probabilities $\{ q_m \}_{m=1}^M$.
Let $c_m \coloneqq q_m \Lambda_m \in \Pos(\V,\W)$; then,
the maximum success probability is equal to the optimal value of
the following SDP problem \cite{Chi-2012-sm,Jen-Pla-2016-sm}:
\begin{alignat}{1}
 \begin{array}{ll}
  \mbox{maximize} & \displaystyle \summ \Tr(\C_{c_m} \Phi_m) \\
  \mbox{subject~to} & \displaystyle \summ \Phi_m = \I_\W \ot \phi \\
 \end{array}
 \label{prob:Phi}
\end{alignat}
with $\Phi_1,\dots,\Phi_M \in \Pos_{\W \ot \V}$
and $\phi \in \Den_\V$.
Note that $\Tr(\C_{c_m} \Phi_k)$ is the joint probability that
the given channel is $\Lambda_m$ and the measurement outcome is $k$.

In what follows, let us consider the problem of discriminating
$\cE_1,\dots,\cE_M$ with prior probabilities $\{ p_m \}_{m=1}^M$.
Assume that each $\cE_m$ is expressed by Eq.~\eqref{eq:cE_Lambda}.
We will present an adaptive discrimination strategy based on Bayesian updating
(see Fig.~\ref{fig:bound_lower_Bayes}),
whose success probability gives a lower bound on the ultimate performance.
As described in the main paper, we will optimize discrimination sequentially for $t = 1, 2, \dots$.

First, in the case of $t = 1$,
we consider the problem of discriminating the channels $\{ \Trp{\W'_1} \Lambda^{(1)}_m \}_{m=1}^M$
with the prior probabilities $\{ p_m \}$.
Note that the partial trace over $\W'_1$ can be regarded as discarding the system $\W'_1$.
This problem is formulated as Problem~\eqref{prob:Phi} with $c_m = p_m \Trp{\W'_1} \Lambda^{(1)}_m$.
Let $\Phi^{(1)} \coloneqq \{ \Phi^{(1)}_m \}_{m=1}^M$ be its optimal solution.
The joint probability that the given process is $\cE_m$ and
the measurement outcome is $k$ is
\begin{alignat}{1}
 q^{(1)}_{m,k} &\coloneqq p_m \Tr \left[ \C_{\Lambda^{(1)}_m} \Phi^{(1)}_k \right].
 \label{eq:q1mk}
\end{alignat}
In this case, the state of the system $\W'_1$ is
\begin{alignat}{1}
 \rho^{(1)}_{m,k} &\coloneqq \frac{1}{q^{(1)}_{m,k}} \cdot p_m
 \Trp{\W_1 \ot \V_1} \left[ \C_{\Lambda^{(1)}_m} \Phi^{(1)}_k \right] \in \Den_{\W'_1}.
 \label{eq:rho1mk}
\end{alignat}
Let $[\varrho^{(1)}, \Pi^{(1)}]$ be
the discrimination associated with $\Phi^{(1)}$;
then, since $\Trp{\W_1 \ot \V_1} \left[ \C_{\Lambda^{(1)}_m} \Phi^{(1)}_k \right] =
\Pi^{(1)}_k \c \Lambda^{(1)}_m \c \varrho^{(1)} \eqqcolon \Gamma_{k|m}^{(1)}$ holds,
we have \cite{Chi-2012-sm,Jen-Pla-2016-sm}
\begin{alignat}{3}
 \trho^{(1)}_{m,k} &\coloneqq q^{(1)}_{m,k} \rho^{(1)}_{m,k} &&= p_m \Gamma_{k|m}^{(1)}
  \in \Pos_{\W'_1}.
\end{alignat}
$\trho^{(1)}_{m,k}$ can be diagrammatically depicted as
\begin{alignat}{1}
 \InsertPDF{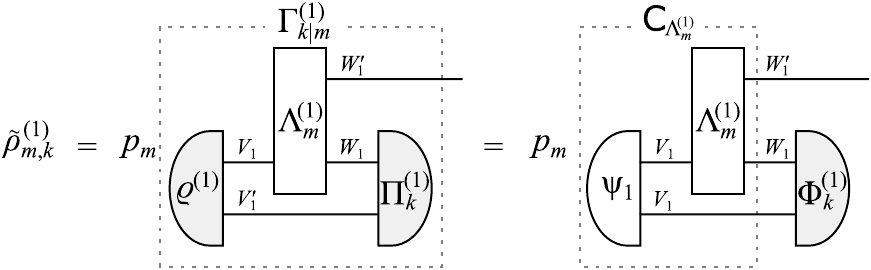} ~\raisebox{1em}{,}
\end{alignat}
where the last equality follows from $\C_{\Lambda^{(1)}_m} = \Lambda^{(1)}_m \c \Psi_1$.
Note that the discrimination $[\varrho^{(1)},\Pi^{(1)}]$ is obtained from $\Phi^{(1)}$
(but is not uniquely determined in general);
however, to obtain the proposed lower bound, we only need $q^{(1)}_{m,k}$ and $\rho^{(1)}_{m,k}$,
which are directly derived from $\Phi^{(1)}$ as in Eqs.~\eqref{eq:q1mk} and \eqref{eq:rho1mk}.

Next, we consider the case of $t = \tau \in \{ 2,\dots,T \}$.
Let $q^{(\tau)}_{m,k}$ be the joint probability that the given process is $\cE_m$ and
the outcome of $\Pi^{(\tau)}$ is $k$
and $\rho^{(\tau)}_{m,k} \in \Den_{\W'_\tau}$ be the state of the system $\W'_\tau$ in this case.
$q^{(\tau)}_{m,k}$ and $\rho^{(\tau)}_{m,k}$ are uniquely determined by
$\trho^{(\tau)}_{m,k} \coloneqq q^{(\tau)}_{m,k} \rho^{(\tau)}_{m,k} \in \Pos_{\W'_\tau}$.
Assume that $\trho^{(\tau-1)}_{m,k}$ is fixed.
Also, assume that $[\varrho^{(\tau)},\Pi^{(\tau)}]$ may depend only on
the outcome of the previous measurement $\Pi^{(\tau-1)}$.
In what follows, the discrimination $[\varrho^{(\tau)},\Pi^{(\tau)}]$ in the case in which
the outcome of the measurement $\Pi^{(\tau-1)}$ is $k_{\tau-1}$ will be often denoted by
$[\varrho^{(\tau)}_{k_{\tau-1}}, \{ \Pi^{(\tau)}_{k|k_{\tau-1}} \}_{k=1}^M]$.
Then, $\trho^{(\tau)}_{m,k}$ is given by
\begin{alignat}{3}
 \trho^{(\tau)}_{m,k}
 &= \sum_{k_{\tau-1}=1}^M \Gamma^{(\tau)}_{k|m,k_{\tau-1}} \c \trho^{(\tau-1)}_{m,k_{\tau-1}},
 \label{eq:trho_taum}
\end{alignat}
where
\begin{alignat}{1}
 \Gamma_{k_t|m,k_{t-1}}^{(t)} &\coloneqq
 \Pi^{(t)}_{k_t|k_{t-1}} \c \Lambda^{(t)}_m \c \varrho^{(t)}_{k_{t-1}}.
\end{alignat}
This can be diagrammatically depicted as
\begin{alignat}{1}
 \InsertPDF{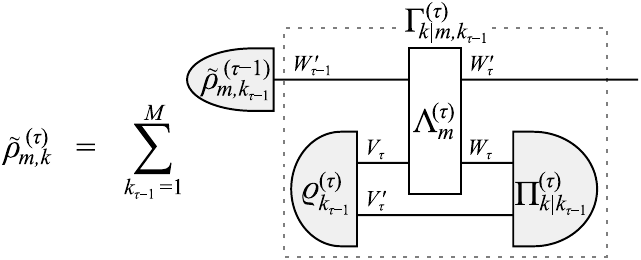} ~\raisebox{1em}{.}
\end{alignat}
The probability that the measurement $\Pi^{(\tau)}$ correctly distinguishes between the processes
$\{ \cE_m \}_{m=1}^M$ is given by
\begin{alignat}{1}
 \PS^{(\tau)} &\coloneqq \summ q^{(\tau)}_{m,m} = \summ \Tr \trho^{(\tau)}_{m,m}.
 \label{eq:PStau}
\end{alignat}
Substituting Eq.~\eqref{eq:trho_taum} into Eq.~\eqref{eq:PStau} yields
\begin{alignat}{2}
 \PS^{(\tau)} &= \sum_{k_{\tau-1}=1}^M R_{k_{\tau-1}}^{(\tau)}, &\quad
 R_{k_{\tau-1}}^{(\tau)} &\coloneqq \summ \Tr
 \left[ \Gamma_{m|m,k_{\tau-1}}^{(\tau)} \c \trho^{(\tau-1)}_{m,k_{\tau-1}} \right].
 \label{eq:PSRk}
\end{alignat}
It is obvious that, for each $k_{\tau-1}$, the larger $R_{k_{\tau-1}}^{(\tau)}$ is,
the larger $\PS^{(\tau)}$ is.
So, we consider finding $[\varrho^{(\tau)}_{k_{\tau-1}}, \{ \Pi^{(\tau)}_{k|{k_{\tau-1}}} \}_{k=1}^M]$
that maximizes $R_{k_{\tau-1}}^{(\tau)}$ for each $k_{\tau-1}$.
This optimization problem is equivalent to Problem~\eqref{prob:Phi}
with $c_m = \Trp{\W'_\tau} \Lambda^{(\tau)}_m \c \trho^{(\tau-1)}_{m,k_{\tau-1}}$.
Let $\{ \Phi^{(\tau)}_{m|k_{\tau-1}} \}_{m=1}^M$ be its optimal solution
[which is associated with the discrimination
$[\varrho^{(\tau)}_{k_{\tau-1}}, \{ \Pi^{(\tau)}_{m|k_{\tau-1}} \}_{m=1}^M]$].
$\trho^{(\tau)}_{m,k}$ is also expressed by
\begin{alignat}{3}
 \trho^{(\tau)}_{m,k} &= \sum_{k_{\tau-1}=1}^M \Trp{\W_\tau \ot \V_\tau}
 \left[ \C_{\Lambda^{(\tau)}_m \c \trho^{(\tau-1)}_{m,k_{\tau-1}}}
 \Phi^{(\tau)}_{k|k_{\tau-1}} \right],
\end{alignat}
or diagrammatically,
\begin{alignat}{1}
 \InsertPDF{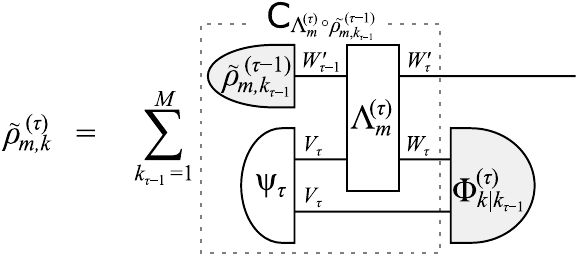} ~\raisebox{1em}{.} 
\end{alignat}
Note that, applying Eq.~\eqref{eq:trho_taum} recursively, we obtain
\begin{alignat}{1}
 \PS^{(\tau)} &= \summ \sum_{k_{\tau-1}=1}^M \sum_{k_{\tau-2}=1}^M \cdots \sum_{k_1=1}^M
 \Tr \left[ \Gamma_{m|m,k_{\tau-1}}^{(\tau)} \c
 \Gamma^{(\tau-1)}_{k_{\tau-1}|m,k_{\tau-2}} \c \cdots \c \Gamma^{(2)}_{k_2|m,k_1} \c \trho^{(1)}_{m,k_1}
 \right],
\end{alignat}
or diagrammatically,
\begin{alignat}{1}
 \InsertPDF{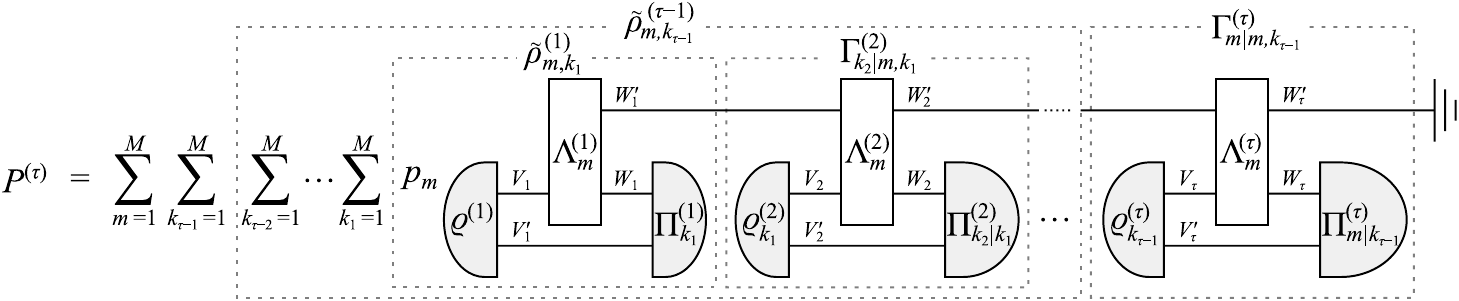} ~\raisebox{1em}{,} 
\end{alignat}
where ``$\gdis$'' denotes the trace.
$\{ \trho^{(2)}_{m,k} \}, \dots, \{ \trho^{(T)}_{m,k} \}$ are sequentially computed
by this algorithm.
The success probability of the discrimination
$[\varrho^{(1)}, \Pi^{(1)}], [\varrho^{(2)}, \Pi^{(2)}], \dots, [\varrho^{(T)}, \Pi^{(T)}]$
is given by
\begin{alignat}{1}
 \lPS \coloneqq P^{(T)}.
\end{alignat}
The ultimate success probability is obviously lower bounded by $\lPS$.

Note that Eq.~\eqref{eq:PSRk} can also be expressed as
\begin{alignat}{2}
 \PS^{(\tau)} &= \sumk q^{(\tau-1)}_k \PS_k^{(\tau)}, &\quad
 \PS_k^{(\tau)} &\coloneqq \summ q^{(\tau-1)}_{m|k} \Tr \left[ \Gamma_{m|m,k}^{(\tau)} \c
 \rho^{(\tau-1)}_{m,k} \right],
 \label{eq:lower_PStau}
\end{alignat}
where $q^{(\tau-1)}_k \coloneqq \summ q^{(\tau-1)}_{m,k}$
is the probability that the outcome of $\Pi^{(\tau-1)}$ is $k$
and $q^{(\tau-1)}_{m|k} \coloneqq q^{(\tau-1)}_{m,k} / q^{(\tau-1)}_k$ is
the conditional probability that the given process is $\cE_m$ given that
the measurement outcome of $\Pi^{(\tau-1)}$ is $k$.
$q^{(\tau-1)}_{m|k}$ can be interpreted as the posterior probability of the process $\cE_m$
given that the outcome of $\Pi^{(\tau-1)}$ is $k$.

We here comment on the computational complexity of obtaining $\lPS$.
It follows that, in the cases of $t = 1$ and $t \ge 2$, respectively, we have to solve
a single-shot channel discrimination problem once and $M$ times,
and $1 + (T-1)M$ times in total.
Thus, the computational complexity is roughly proportional to $TM$ for relatively large $T$.

A similar argument used to derive Eq.~\eqref{eq:bound_upper2}
can be easily extended to obtain a tighter lower bound,
at the expense of additional complexity.
For instance, instead of considering (single-shot) discrimination of channels
$\{ \Lambda^{(t)}_m \}_{m=1}^M$,
we can consider discrimination of processes $\{ \Lambda^{(t)}_m \ast \Lambda^{(t-1)}_m \}_{m=1}^M$.

\section{Discrimination of memory channels associated with two consecutive uses of
generalized AD channels with correlated noise} \label{supplemental:numericalGAD}

A memory channel, $\GAD$, associated with two consecutive uses of generalized AD channel
with correlated noise is characterized by the three parameters, $\pc$, $\nu$, and $n$, and is defined as
\begin{alignat}{1}
 \GAD(\rho) &\coloneqq (1 - \pc) \GAD_\mathrm{u}(\rho) + \pc \GAD_\mathrm{c}(\rho),
 \nonumber \\
 \GAD_\mathrm{u}(\rho) &\coloneqq \sum_{j=1}^4 \sum_{k=1}^4 (E_j \ot E_k) \rho (E_j \ot E_k)^\dagger,
 \nonumber \\
 \GAD_\mathrm{c}(\rho) &\coloneqq \sum_{j=1}^5 B_j \rho B_j^\dagger.
 \label{eq:GAD}
\end{alignat}
$E_j$ is
\begin{alignat}{1}
 E_1 &\coloneqq \sqrt{\gamma}
 \begin{bmatrix}
  1 & 0 \\
  0 & \sqrt{\omega} \\
 \end{bmatrix}, \quad
 E_2 \coloneqq \sqrt{1 - \gamma}
 \begin{bmatrix}
  \sqrt{\omega} & 0 \\
  0 & 1 \\
 \end{bmatrix}, \quad
 E_3 \coloneqq \sqrt{\gamma}
 \begin{bmatrix}
  0 & \sqrt{1 - \omega} \\
  0 & 0 \\
 \end{bmatrix}, \quad
 E_4 \coloneqq \sqrt{1 - \gamma}
 \begin{bmatrix}
  0 & 0 \\
  \sqrt{1 - \omega} & 0 \\
 \end{bmatrix},
\end{alignat}
where $\gamma \coloneqq \frac{n}{2n+1}$ and $\omega \coloneqq e^{-(2n+1)\nu}$.
$B_j$ is
\begin{alignat}{3}
 B_1 &\coloneqq
 \begin{bmatrix}
  \sqrt{e^{-(n+1)\nu}} & 0 & 0 & 0 \\
  0 & 1 & 0 & 0 \\
  0 & 0 & 1 & 0 \\
  0 & 0 & 0 & \sqrt{e^{-n\nu}} \\
 \end{bmatrix}, &\quad
 B_2 &\coloneqq
 \begin{bmatrix}
  0 & 0 & 0 & 0 \\
  0 & 0 & 0 & 0 \\
  0 & 0 & 0 & 0 \\
  \sqrt{(1-\gamma)(1-\omega)} & 0 & 0 & 0 \\
 \end{bmatrix}, &\quad
 B_3 &\coloneqq
 \begin{bmatrix}
  0 & 0 & 0 & \sqrt{\gamma(1-\omega)} \\
  0 & 0 & 0 & 0 \\
  0 & 0 & 0 & 0 \\
  0 & 0 & 0 & 0 \\
 \end{bmatrix}, \nonumber \\
 B_4 &\coloneqq
 \begin{bmatrix}
  \sqrt{\gamma + \omega - \gamma\omega - e^{-(n+1)\nu}} & 0 & 0 & 0 \\
  0 & 0 & 0 & 0 \\
  0 & 0 & 0 & 0 \\
  0 & 0 & 0 & 0 \\
 \end{bmatrix}, &\quad
 B_5 &\coloneqq
 \begin{bmatrix}
  0 & 0 & 0 & 0 \\
  0 & 0 & 0 & 0 \\
  0 & 0 & 0 & 0 \\
  0 & 0 & 0 & \sqrt{1 - \gamma + \gamma\omega - e^{-n\nu}} \\
 \end{bmatrix}.
\end{alignat}
Equation~\eqref{eq:GAD} is obtained by substituting $m = 0$ into Eq.~(6) of Ref.~\cite{Jeo-Shi-2019-sm}.
(The term under the square root sign in this equation is negative in some cases,
so we slightly modify it.)
Note that $\pc$ is the degree of memory,
$\nu$ is associated with the zero-temperature dissipation rate,
and $n$ is associated with the number of thermal photons.

In the numerical experiment of Fig.~\ref{fig:result-GAD_memory},
we considered the problem of discriminating three processes $\cE_1$, $\cE_2$, and $\cE_3$
with equal prior probabilities, where
\begin{alignat}{1}
 \cE_m &\coloneqq \left( \Trp{\W'} \GAD_{\delta_{m,3}} \right) \ast \GAD_{\delta_{m,2}}
 \ast \left( \GAD_{\delta_{m,1}} \c \sigma_\mathrm{ex} \right)
\end{alignat}
and $\sigma_\mathrm{ex} \coloneqq \ket{0}\bra{0} \in \Den_{\W'}$.
$\GAD_r \in \Chn(\W' \ot \V, \W' \ot \W)$ $~(r \in \{0,1\})$ is
a memory channel associated with two consecutive uses of generalized AD channel
with correlated noise, where $\V$, $\W$, and $\W'$ are qubit systems.
The conditional probability that the measurement outcome is $k$ given that the given process is $\cE_m$
is diagrammatically depicted as
\begin{alignat}{1}
 \InsertPDF{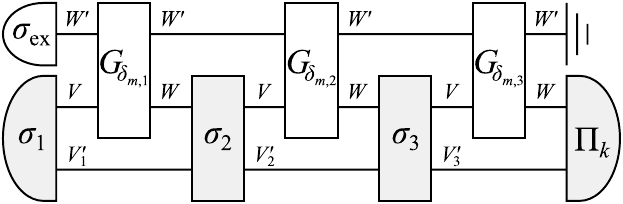} ~\raisebox{.1em}{.}
\end{alignat}
We set $\pc = 0.2$ and $n = 1$ for both $\GAD_0$ and $\GAD_1$.
In Fig.~\ref{fig:result-GAD_memory},
we plot the success probability as a function of the parameter $\nu$ of $\GAD_0$, $\nu_0$,
where the parameter $\nu$ of $\GAD_1$ is set to $\nu_0 + 0.04$.

\RA{
\section{Computational complexity of obtaining upper and lower bounds} \label{supplemental:complexity}
}

\RA{We study the computational complexity of computing proposed upper and lower bounds.
Let us consider the problem of discriminating $M$ processes $\cE_1,\dots,\cE_M$,
where each $\cE_m$ is a process consisting of $T$ channels $\Lambda^{(1)}_m,\dots,\Lambda^{(T)}_m$
as in Fig.~\ref{fig:process_discrimination}.
Let $n_t$ and $n^{(2)}_{2t}$ be, respectively, the order of the Choi-Jamio{\l}kowski representation
of $\Lambda^{(t)}_m$ and $\Lambda^{(2t)}_m \ast \Lambda^{(2t-1)}_m$.
One can easily verify
\begin{alignat}{1}
 n_t = N_{\W'_t \ot \W_t \ot \W'_{t-1} \ot \V_t}, \quad
 n^{(2)}_{2t} = N_{\W'_{2t} \ot \W_{2t} \ot \V_{2t} \ot \W_{2t-1} \ot \W'_{2t-2} \ot \V_{2t-1}}.
\end{alignat}
Also, let $\mT(L)$ denote the time required to solve an SDP problem of order $L$.
Note that $\mT(L) = \poly(L)$ holds; a typical SDP solver satisfies $\mT(L) = O(L^6)$ \cite{Bor-1999-sm}.

The total time complexities required to obtain
$\PS^\opt$, $\oPS{1}$, $\oPS{2}$, and $\lPS$ are summarized in Table~\ref{table:complexity}.
We can compute $\PS^\opt$ by solving an SDP problem of order
$\prod_{t=1}^T N_{\W_t \ot \V_t}$, which takes $\mT(\prod_{t=1}^T N_{\W_t \ot \V_t})$ time.
We here focus on the even $T$ case since the odd case is similar.
To obtain $\oPS{1}$ (resp. $\oPS{2}$), we need to solve an SDP problem of order
$n_t$ [resp. $n^{(2)}_{2t}$] for each $t \in \{1,2,\dots,T\}$ (resp. $t \in \{1,2,\dots,T/2\}$),
and thus the total time complexity is $\sum_{t=1}^T \mT(n_t)$
[resp. $\sum_{t=1}^{T/2} \mT[n^{(2)}_{2t}]$].
To obtain $\lPS$, we need to solve an SDP problem of order $n_1$
and, for each $t \in \{2,\dots,T\}$, $M$ SDP problems of order $n_t$,
which yields a total time complexity of $\mT(n_1) + M \sum_{t=2}^T \mT(n_t)$.
}%
\begin{table}[btp]
 \caption{\RA{Time complexity for computing the ultimate success probability and its proposed upper and lower bounds.}}
 \label{table:complexity}
 \begin{center}
  \begingroup
  \RA{
  \renewcommand{\arraystretch}{1.2}
  \begin{tabular}{|c|c|}
   \hline
   Value & Time complexity \\ \hline
   $\PS^\opt$ & $\mT(\prod_{t=1}^T N_{\W_t \ot \V_t})$ \\ \hline
   $\oPS{1}$ & $\sum_{t=1}^T \mT(n_t)$ \\ \hline
   $\oPS{2}$ (for even $T$) & $\sum_{t=1}^{T/2} \mT[n^{(2)}_{2t}]$ \\ \hline
   $\lPS$ & $\mT(n_1) + M \sum_{t=2}^T \mT(n_t)$ \\ \hline
  \end{tabular}}
  \endgroup
 \end{center}
\end{table}

\RA{
We now consider the numerical experiment of Fig.~\ref{fig:result-CPF},
which handles the problem of channel position finding with AD channels \cite{Zhu-Pir-2020-sm},
i.e., the problem of $T$-shot discrimination of $M$ channels $\{ \tLambda_m \}_{m=1}^M$
defined by Eq.~\eqref{eq:CPF_Lambda}.
We can easily check  $N_{\W'_t} = 1$, $N_{\W_t} = N_{\V_t} = 2^M$,
$n_t = 4^M$, and $n^{(2)}_{2t} = 16^M$.
This gives that the time complexities required for computing
$\PS^\opt$, $\oPS{1}$, $\oPS{2}$, and $\lPS$ are,
respectively, $\mT(4^{MT})$, $\mT(4^M)$, $\mT(16^M)$, and $[1 + M(T-1)]\mT(4^M) = O(TM)\mT(4^M)$.
Note that, in this problem, since $\Lambda^{(t)}_m$ of Fig.~\ref{fig:process_discrimination}
is independent of $t$,
we can compute $\oPS{1}$ (resp. $\oPS{2}$) by solving only one SDP problem of order
$n_1$ [resp. $n^{(2)}_2$].
The time complexity required for computing $\ol{P'_1}$ is $O(1)$,
in which we need to solve only one SDP problem of Eq.~\eqref{prob:sAD2},
which can be solved analytically as described in Sec.~\ref{supplemental:CPFAD}.

We then discuss the numerical experiment of Fig.~\ref{fig:result-GAD_memory},
in which case $T = M$ holds.
We can easily check $N_{\W_t} = N_{\V_t} = N_{\W'_t} = 2$, and $n_1 = 8$.
Thus, the time complexities required for computing $\PS^\opt$, $\oPS{1}$, $\oPS{2}$,
and $\lPS$ are, respectively,
$\mT(4^M)$, $O(1)$, $O(1)$, and $[1 + M(M-1)]\mT(4) = O(M^2)$.
Note that we can compute each of $\oPS{1}$ and $\oPS{2}$ by solving at most three SDP problems
of fixed orders.
Note that we can compute $\oPS{1}$ by solving only three SDP problems of the fixed orders
$n_1 = 8$, $n_2 = 16$, and $n_M = 8$.
The same argument holds for computing $\oPS{2}$.}


\begin{thebibliography}{71}%
\makeatletter
\providecommand \@ifxundefined [1]{%
 \@ifx{#1\undefined}
}%
\providecommand \@ifnum [1]{%
 \ifnum #1\expandafter \@firstoftwo
 \else \expandafter \@secondoftwo
 \fi
}%
\providecommand \@ifx [1]{%
 \ifx #1\expandafter \@firstoftwo
 \else \expandafter \@secondoftwo
 \fi
}%
\providecommand \natexlab [1]{#1}%
\providecommand \enquote  [1]{``#1''}%
\providecommand \bibnamefont  [1]{#1}%
\providecommand \bibfnamefont [1]{#1}%
\providecommand \citenamefont [1]{#1}%
\providecommand \href@noop [0]{\@secondoftwo}%
\providecommand \href [0]{\begingroup \@sanitize@url \@href}%
\providecommand \@href[1]{\@@startlink{#1}\@@href}%
\providecommand \@@href[1]{\endgroup#1\@@endlink}%
\providecommand \@sanitize@url [0]{\catcode `\\12\catcode `\$12\catcode
  `\&12\catcode `\#12\catcode `\^12\catcode `\_12\catcode `\%12\relax}%
\providecommand \@@startlink[1]{}%
\providecommand \@@endlink[0]{}%
\providecommand \url  [0]{\begingroup\@sanitize@url \@url }%
\providecommand \@url [1]{\endgroup\@href {#1}{\urlprefix }}%
\providecommand \urlprefix  [0]{URL }%
\providecommand \Eprint [0]{\href }%
\providecommand \doibase [0]{http://dx.doi.org/}%
\providecommand \selectlanguage [0]{\@gobble}%
\providecommand \bibinfo  [0]{\@secondoftwo}%
\providecommand \bibfield  [0]{\@secondoftwo}%
\providecommand \translation [1]{[#1]}%
\providecommand \BibitemOpen [0]{}%
\providecommand \bibitemStop [0]{}%
\providecommand \bibitemNoStop [0]{.\EOS\space}%
\providecommand \EOS [0]{\spacefactor3000\relax}%
\providecommand \BibitemShut  [1]{\csname bibitem#1\endcsname}%
\let\auto@bib@innerbib\@empty
%</preamble>
\bibitem [{\citenamefont {Helstrom}(1969)}]{Hel-1969}%
  \BibitemOpen
  \bibfield  {author} {\bibinfo {author} {\bibfnamefont {C.~W.}\ \bibnamefont
  {Helstrom}},\ }\href {https://doi.org/10.1007/BF01007479} {\bibfield
  {journal} {\bibinfo  {journal} {J. Stat. Phys.}\ }\textbf {\bibinfo {volume}
  {1}},\ \bibinfo {pages} {231} (\bibinfo {year} {1969})}\BibitemShut {NoStop}%
\bibitem [{\citenamefont {Holevo}(1973)}]{Hol-1973}%
  \BibitemOpen
  \bibfield  {author} {\bibinfo {author} {\bibfnamefont {A.~S.}\ \bibnamefont
  {Holevo}},\ }\href {https://doi.org/10.1016/0047-259X(73)90028-6} {\bibfield
  {journal} {\bibinfo  {journal} {J. Multivar. Anal.}\ }\textbf {\bibinfo
  {volume} {3}},\ \bibinfo {pages} {337} (\bibinfo {year} {1973})}\BibitemShut
  {NoStop}%
\bibitem [{\citenamefont {Yuen}\ \emph {et~al.}(1975)\citenamefont {Yuen},
  \citenamefont {Kennedy},\ and\ \citenamefont {Lax}}]{Yue-Ken-Lax-1975}%
  \BibitemOpen
  \bibfield  {author} {\bibinfo {author} {\bibfnamefont {H.~P.}\ \bibnamefont
  {Yuen}}, \bibinfo {author} {\bibfnamefont {K.~S.}\ \bibnamefont {Kennedy}}, \
  and\ \bibinfo {author} {\bibfnamefont {M.}~\bibnamefont {Lax}},\ }\href
  {https://doi.org/10.1109/TIT.1975.1055351} {\bibfield  {journal} {\bibinfo
  {journal} {IEEE Trans. Inf. Theory}\ }\textbf {\bibinfo {volume} {21}},\
  \bibinfo {pages} {125} (\bibinfo {year} {1975})}\BibitemShut {NoStop}%
\bibitem [{\citenamefont {Belavkin}(1975)}]{Bel-1975}%
  \BibitemOpen
  \bibfield  {author} {\bibinfo {author} {\bibfnamefont {V.~P.}\ \bibnamefont
  {Belavkin}},\ }\href {https://doi.org/10.1080/17442507508833114} {\bibfield
  {journal} {\bibinfo  {journal} {Stochastics}\ }\textbf {\bibinfo {volume}
  {1}},\ \bibinfo {pages} {315} (\bibinfo {year} {1975})}\BibitemShut {NoStop}%
\bibitem [{\citenamefont {Hayashi}\ \emph {et~al.}(2008)\citenamefont
  {Hayashi}, \citenamefont {Kawachi},\ and\ \citenamefont
  {Kobayashi}}]{Hay-Kaw-Kob-2008}%
  \BibitemOpen
  \bibfield  {author} {\bibinfo {author} {\bibfnamefont {M.}~\bibnamefont
  {Hayashi}}, \bibinfo {author} {\bibfnamefont {A.}~\bibnamefont {Kawachi}}, \
  and\ \bibinfo {author} {\bibfnamefont {H.}~\bibnamefont {Kobayashi}},\ }\href
  {https://doi.org/10.26421/QIC8.3-4-8} {\bibfield  {journal} {\bibinfo
  {journal} {Quantum Inf. Comput.}\ }\textbf {\bibinfo {volume} {8}},\ \bibinfo
  {pages} {0345} (\bibinfo {year} {2008})}\BibitemShut {NoStop}%
\bibitem [{\citenamefont {Montanaro}(2008)}]{Mon-2008}%
  \BibitemOpen
  \bibfield  {author} {\bibinfo {author} {\bibfnamefont {A.}~\bibnamefont
  {Montanaro}},\ }in\ \href {https://doi.org/10.1109/ITW.2008.4578690} {\emph
  {\bibinfo {booktitle} {2008 IEEE Information Theory Workshop}}}\ (\bibinfo
  {organization} {IEEE},\ \bibinfo {year} {2008})\ pp.\ \bibinfo {pages}
  {378--380}\BibitemShut {NoStop}%
\bibitem [{\citenamefont {Qiu}(2008)}]{Qiu-2008}%
  \BibitemOpen
  \bibfield  {author} {\bibinfo {author} {\bibfnamefont {D.}~\bibnamefont
  {Qiu}},\ }\href {https://doi.org/10.1103/PhysRevA.77.012328} {\bibfield
  {journal} {\bibinfo  {journal} {Phys. Rev. A}\ }\textbf {\bibinfo {volume}
  {77}},\ \bibinfo {pages} {012328} (\bibinfo {year} {2008})}\BibitemShut
  {NoStop}%
\bibitem [{\citenamefont {Tyson}(2009)}]{Tys-2009}%
  \BibitemOpen
  \bibfield  {author} {\bibinfo {author} {\bibfnamefont {J.}~\bibnamefont
  {Tyson}},\ }\href {https://doi.org/10.1103/PhysRevA.79.032343} {\bibfield
  {journal} {\bibinfo  {journal} {Phys. Rev. A}\ }\textbf {\bibinfo {volume}
  {79}},\ \bibinfo {pages} {032343} (\bibinfo {year} {2009})}\BibitemShut
  {NoStop}%
\bibitem [{\citenamefont {Qiu}\ and\ \citenamefont {Li}(2010)}]{Qiu-Li-2010}%
  \BibitemOpen
  \bibfield  {author} {\bibinfo {author} {\bibfnamefont {D.}~\bibnamefont
  {Qiu}}\ and\ \bibinfo {author} {\bibfnamefont {L.}~\bibnamefont {Li}},\
  }\href {https://doi.org/10.1103/PhysRevA.81.042329} {\bibfield  {journal}
  {\bibinfo  {journal} {Phys. Rev. A}\ }\textbf {\bibinfo {volume} {81}},\
  \bibinfo {pages} {042329} (\bibinfo {year} {2010})}\BibitemShut {NoStop}%
\bibitem [{\citenamefont {Ji}\ \emph {et~al.}(2006)\citenamefont {Ji},
  \citenamefont {Feng}, \citenamefont {Duan},\ and\ \citenamefont
  {Ying}}]{Ji-Fen-Dua-Yin-2006}%
  \BibitemOpen
  \bibfield  {author} {\bibinfo {author} {\bibfnamefont {Z.}~\bibnamefont
  {Ji}}, \bibinfo {author} {\bibfnamefont {Y.}~\bibnamefont {Feng}}, \bibinfo
  {author} {\bibfnamefont {R.}~\bibnamefont {Duan}}, \ and\ \bibinfo {author}
  {\bibfnamefont {M.}~\bibnamefont {Ying}},\ }\href
  {https://doi.org/10.1103/PhysRevLett.96.200401} {\bibfield  {journal}
  {\bibinfo  {journal} {Phys. Rev. Lett.}\ }\textbf {\bibinfo {volume} {96}},\
  \bibinfo {pages} {200401} (\bibinfo {year} {2006})}\BibitemShut {NoStop}%
\bibitem [{\citenamefont {Ziman}\ and\ \citenamefont
  {Heinosaari}(2008)}]{Zim-Hei-2008}%
  \BibitemOpen
  \bibfield  {author} {\bibinfo {author} {\bibfnamefont {M.}~\bibnamefont
  {Ziman}}\ and\ \bibinfo {author} {\bibfnamefont {T.}~\bibnamefont
  {Heinosaari}},\ }\href {https://doi.org/10.1103/PhysRevA.77.042321}
  {\bibfield  {journal} {\bibinfo  {journal} {Phys. Rev. A}\ }\textbf {\bibinfo
  {volume} {77}},\ \bibinfo {pages} {042321} (\bibinfo {year}
  {2008})}\BibitemShut {NoStop}%
\bibitem [{\citenamefont {Ziman}\ \emph {et~al.}(2009)\citenamefont {Ziman},
  \citenamefont {Heinosaari},\ and\ \citenamefont
  {Sedl{\'a}k}}]{Zim-Hei-Sed-2009}%
  \BibitemOpen
  \bibfield  {author} {\bibinfo {author} {\bibfnamefont {M.}~\bibnamefont
  {Ziman}}, \bibinfo {author} {\bibfnamefont {T.}~\bibnamefont {Heinosaari}}, \
  and\ \bibinfo {author} {\bibfnamefont {M.}~\bibnamefont {Sedl{\'a}k}},\
  }\href {https://doi.org/10.1103/PhysRevA.80.052102} {\bibfield  {journal}
  {\bibinfo  {journal} {Phys. Rev. A}\ }\textbf {\bibinfo {volume} {80}},\
  \bibinfo {pages} {052102} (\bibinfo {year} {2009})}\BibitemShut {NoStop}%
\bibitem [{\citenamefont {Sedl{\'a}k}\ and\ \citenamefont
  {Ziman}(2014)}]{Sed-Zim-2014}%
  \BibitemOpen
  \bibfield  {author} {\bibinfo {author} {\bibfnamefont {M.}~\bibnamefont
  {Sedl{\'a}k}}\ and\ \bibinfo {author} {\bibfnamefont {M.}~\bibnamefont
  {Ziman}},\ }\href {https://doi.org/10.1103/PhysRevA.90.052312} {\bibfield
  {journal} {\bibinfo  {journal} {Phys. Rev. A}\ }\textbf {\bibinfo {volume}
  {90}},\ \bibinfo {pages} {052312} (\bibinfo {year} {2014})}\BibitemShut
  {NoStop}%
\bibitem [{\citenamefont {Pucha{\l}a}\ \emph {et~al.}(2018)\citenamefont
  {Pucha{\l}a}, \citenamefont {Pawela}, \citenamefont {Krawiec},\ and\
  \citenamefont {Kukulski}}]{Puc-Paw-Kra-Kuk-2018}%
  \BibitemOpen
  \bibfield  {author} {\bibinfo {author} {\bibfnamefont {Z.}~\bibnamefont
  {Pucha{\l}a}}, \bibinfo {author} {\bibfnamefont {{\L}.}~\bibnamefont
  {Pawela}}, \bibinfo {author} {\bibfnamefont {A.}~\bibnamefont {Krawiec}}, \
  and\ \bibinfo {author} {\bibfnamefont {R.}~\bibnamefont {Kukulski}},\ }\href
  {https://doi.org/10.1103/PhysRevA.98.042103} {\bibfield  {journal} {\bibinfo
  {journal} {Phys. Rev. A}\ }\textbf {\bibinfo {volume} {98}},\ \bibinfo
  {pages} {042103} (\bibinfo {year} {2018})}\BibitemShut {NoStop}%
\bibitem [{\citenamefont {Krawiec}\ \emph {et~al.}(2020)\citenamefont
  {Krawiec}, \citenamefont {Pawela},\ and\ \citenamefont
  {Pucha{\l}a}}]{Kra-Paw-Puc-2020}%
  \BibitemOpen
  \bibfield  {author} {\bibinfo {author} {\bibfnamefont {A.}~\bibnamefont
  {Krawiec}}, \bibinfo {author} {\bibfnamefont {{\L}.}~\bibnamefont {Pawela}},
  \ and\ \bibinfo {author} {\bibfnamefont {Z.}~\bibnamefont {Pucha{\l}a}},\
  }\href {https://doi.org/10.1007/s11128-020-02883-3} {\bibfield  {journal}
  {\bibinfo  {journal} {Quant. Inf. Proc.}\ }\textbf {\bibinfo {volume} {19}},\
  \bibinfo {pages} {428} (\bibinfo {year} {2020})}\BibitemShut {NoStop}%
\bibitem [{\citenamefont {Datta}\ \emph {et~al.}(2021)\citenamefont {Datta},
  \citenamefont {Biswas}, \citenamefont {Saha},\ and\ \citenamefont
  {Augusiak}}]{Dat-Bis-Sah-Aug-2020}%
  \BibitemOpen
  \bibfield  {author} {\bibinfo {author} {\bibfnamefont {C.}~\bibnamefont
  {Datta}}, \bibinfo {author} {\bibfnamefont {T.}~\bibnamefont {Biswas}},
  \bibinfo {author} {\bibfnamefont {D.}~\bibnamefont {Saha}}, \ and\ \bibinfo
  {author} {\bibfnamefont {R.}~\bibnamefont {Augusiak}},\ }\href
  {https://doi.org/10.1088/1367-2630/abecaf} {\bibfield  {journal} {\bibinfo
  {journal} {New J. Phys.}\ }\textbf {\bibinfo {volume} {23}},\ \bibinfo
  {pages} {043021} (\bibinfo {year} {2021})}\BibitemShut {NoStop}%
\bibitem [{\citenamefont {Acin}(2001)}]{Aci-2001}%
  \BibitemOpen
  \bibfield  {author} {\bibinfo {author} {\bibfnamefont {A.}~\bibnamefont
  {Acin}},\ }\href {https://doi.org/10.1103/PhysRevLett.87.177901} {\bibfield
  {journal} {\bibinfo  {journal} {Phys. Rev. Lett.}\ }\textbf {\bibinfo
  {volume} {87}},\ \bibinfo {pages} {177901} (\bibinfo {year}
  {2001})}\BibitemShut {NoStop}%
\bibitem [{\citenamefont {Sacchi}(2005{\natexlab{a}})}]{Sac-2005}%
  \BibitemOpen
  \bibfield  {author} {\bibinfo {author} {\bibfnamefont {M.~F.}\ \bibnamefont
  {Sacchi}},\ }\href {https://doi.org/10.1103/PhysRevA.71.062340} {\bibfield
  {journal} {\bibinfo  {journal} {Phys. Rev. A}\ }\textbf {\bibinfo {volume}
  {71}},\ \bibinfo {pages} {062340} (\bibinfo {year}
  {2005}{\natexlab{a}})}\BibitemShut {NoStop}%
\bibitem [{\citenamefont {Sacchi}(2005{\natexlab{b}})}]{Sac-2005-EB}%
  \BibitemOpen
  \bibfield  {author} {\bibinfo {author} {\bibfnamefont {M.~F.}\ \bibnamefont
  {Sacchi}},\ }\href {https://doi.org/10.1103/PhysRevA.72.014305} {\bibfield
  {journal} {\bibinfo  {journal} {Phys. Rev. A}\ }\textbf {\bibinfo {volume}
  {72}},\ \bibinfo {pages} {014305} (\bibinfo {year}
  {2005}{\natexlab{b}})}\BibitemShut {NoStop}%
\bibitem [{\citenamefont {Li}\ and\ \citenamefont {Qiu}(2008)}]{Li-Qiu-2008}%
  \BibitemOpen
  \bibfield  {author} {\bibinfo {author} {\bibfnamefont {L.}~\bibnamefont
  {Li}}\ and\ \bibinfo {author} {\bibfnamefont {D.}~\bibnamefont {Qiu}},\
  }\href {https://doi.org/10.1088/1751-8113/41/33/335302} {\bibfield  {journal}
  {\bibinfo  {journal} {J. Phys. A: Math. Theor.}\ }\textbf {\bibinfo {volume}
  {41}},\ \bibinfo {pages} {335302} (\bibinfo {year} {2008})}\BibitemShut
  {NoStop}%
\bibitem [{\citenamefont {Pirandola}\ and\ \citenamefont
  {Lupo}(2017)}]{Pir-Lup-2017}%
  \BibitemOpen
  \bibfield  {author} {\bibinfo {author} {\bibfnamefont {S.}~\bibnamefont
  {Pirandola}}\ and\ \bibinfo {author} {\bibfnamefont {C.}~\bibnamefont
  {Lupo}},\ }\href {https://doi.org/10.1103/PhysRevLett.118.100502} {\bibfield
  {journal} {\bibinfo  {journal} {Phys. Rev. Lett.}\ }\textbf {\bibinfo
  {volume} {118}},\ \bibinfo {pages} {100502} (\bibinfo {year}
  {2017})}\BibitemShut {NoStop}%
\bibitem [{\citenamefont {Pirandola}\ \emph {et~al.}(2019)\citenamefont
  {Pirandola}, \citenamefont {Laurenza}, \citenamefont {Lupo},\ and\
  \citenamefont {Pereira}}]{Pir-Lau-Lup-Per-2019}%
  \BibitemOpen
  \bibfield  {author} {\bibinfo {author} {\bibfnamefont {S.}~\bibnamefont
  {Pirandola}}, \bibinfo {author} {\bibfnamefont {R.}~\bibnamefont {Laurenza}},
  \bibinfo {author} {\bibfnamefont {C.}~\bibnamefont {Lupo}}, \ and\ \bibinfo
  {author} {\bibfnamefont {J.~L.}\ \bibnamefont {Pereira}},\ }\href
  {https://doi.org/10.1038/s41534-019-0162-y} {\bibfield  {journal} {\bibinfo
  {journal} {npj Quantum Inf.}\ }\textbf {\bibinfo {volume} {5}},\ \bibinfo
  {pages} {50} (\bibinfo {year} {2019})}\BibitemShut {NoStop}%
\bibitem [{\citenamefont {Chiribella}\ \emph
  {et~al.}(2008{\natexlab{a}})\citenamefont {Chiribella}, \citenamefont
  {D'{A}riano},\ and\ \citenamefont {Perinotti}}]{Chi-Dar-Per-2008-supermap}%
  \BibitemOpen
  \bibfield  {author} {\bibinfo {author} {\bibfnamefont {G.}~\bibnamefont
  {Chiribella}}, \bibinfo {author} {\bibfnamefont {G.~M.}\ \bibnamefont
  {D'{A}riano}}, \ and\ \bibinfo {author} {\bibfnamefont {P.}~\bibnamefont
  {Perinotti}},\ }\href {https://doi.org/10.1209/0295-5075/83/30004} {\bibfield
   {journal} {\bibinfo  {journal} {EPL (Europhysics Letters)}\ }\textbf
  {\bibinfo {volume} {83}},\ \bibinfo {pages} {30004} (\bibinfo {year}
  {2008}{\natexlab{a}})}\BibitemShut {NoStop}%
\bibitem [{\citenamefont {Chiribella}\ \emph
  {et~al.}(2008{\natexlab{b}})\citenamefont {Chiribella}, \citenamefont
  {D'{A}riano},\ and\ \citenamefont {Perinotti}}]{Chi-Dar-Per-2008}%
  \BibitemOpen
  \bibfield  {author} {\bibinfo {author} {\bibfnamefont {G.}~\bibnamefont
  {Chiribella}}, \bibinfo {author} {\bibfnamefont {G.~M.}\ \bibnamefont
  {D'{A}riano}}, \ and\ \bibinfo {author} {\bibfnamefont {P.}~\bibnamefont
  {Perinotti}},\ }\href {https://doi.org/10.1103/PhysRevLett.101.060401}
  {\bibfield  {journal} {\bibinfo  {journal} {Phys. Rev. Lett.}\ }\textbf
  {\bibinfo {volume} {101}},\ \bibinfo {pages} {060401} (\bibinfo {year}
  {2008}{\natexlab{b}})}\BibitemShut {NoStop}%
\bibitem [{\citenamefont {Macchiavello}\ and\ \citenamefont
  {Palma}(2002)}]{Mac-Pal-2002}%
  \BibitemOpen
  \bibfield  {author} {\bibinfo {author} {\bibfnamefont {C.}~\bibnamefont
  {Macchiavello}}\ and\ \bibinfo {author} {\bibfnamefont {G.~M.}\ \bibnamefont
  {Palma}},\ }\href {https://doi.org/10.1103/PhysRevA.65.050301} {\bibfield
  {journal} {\bibinfo  {journal} {Phys. Rev. A}\ }\textbf {\bibinfo {volume}
  {65}},\ \bibinfo {pages} {050301} (\bibinfo {year} {2002})}\BibitemShut
  {NoStop}%
\bibitem [{\citenamefont {Yeo}\ and\ \citenamefont
  {Skeen}(2003)}]{Yeo-Ske-2003}%
  \BibitemOpen
  \bibfield  {author} {\bibinfo {author} {\bibfnamefont {Y.}~\bibnamefont
  {Yeo}}\ and\ \bibinfo {author} {\bibfnamefont {A.}~\bibnamefont {Skeen}},\
  }\href {https://doi.org/10.1103/PhysRevA.67.064301} {\bibfield  {journal}
  {\bibinfo  {journal} {Phys. Rev. A}\ }\textbf {\bibinfo {volume} {67}},\
  \bibinfo {pages} {064301} (\bibinfo {year} {2003})}\BibitemShut {NoStop}%
\bibitem [{\citenamefont {Bowen}\ and\ \citenamefont
  {Mancini}(2004)}]{Bow-Man-2004}%
  \BibitemOpen
  \bibfield  {author} {\bibinfo {author} {\bibfnamefont {G.}~\bibnamefont
  {Bowen}}\ and\ \bibinfo {author} {\bibfnamefont {S.}~\bibnamefont
  {Mancini}},\ }\href {https://doi.org/10.1103/PhysRevA.69.012306} {\bibfield
  {journal} {\bibinfo  {journal} {Phys. Rev. A}\ }\textbf {\bibinfo {volume}
  {69}},\ \bibinfo {pages} {012306} (\bibinfo {year} {2004})}\BibitemShut
  {NoStop}%
\bibitem [{\citenamefont {Kretschmann}\ and\ \citenamefont
  {Werner}(2005)}]{Kre-Wer-2005}%
  \BibitemOpen
  \bibfield  {author} {\bibinfo {author} {\bibfnamefont {D.}~\bibnamefont
  {Kretschmann}}\ and\ \bibinfo {author} {\bibfnamefont {R.~F.}\ \bibnamefont
  {Werner}},\ }\href {https://doi.org/10.1103/PhysRevA.72.062323} {\bibfield
  {journal} {\bibinfo  {journal} {Phys. Rev. A}\ }\textbf {\bibinfo {volume}
  {72}},\ \bibinfo {pages} {062323} (\bibinfo {year} {2005})}\BibitemShut
  {NoStop}%
\bibitem [{\citenamefont {Plenio}\ and\ \citenamefont
  {Virmani}(2007)}]{Ple-Vir-2007}%
  \BibitemOpen
  \bibfield  {author} {\bibinfo {author} {\bibfnamefont {M.~B.}\ \bibnamefont
  {Plenio}}\ and\ \bibinfo {author} {\bibfnamefont {S.}~\bibnamefont
  {Virmani}},\ }\href {https://doi.org/10.1103/PhysRevLett.99.120504}
  {\bibfield  {journal} {\bibinfo  {journal} {Phys. Rev. Lett.}\ }\textbf
  {\bibinfo {volume} {99}},\ \bibinfo {pages} {120504} (\bibinfo {year}
  {2007})}\BibitemShut {NoStop}%
\bibitem [{\citenamefont {Pirandola}(2011)}]{Pir-2011}%
  \BibitemOpen
  \bibfield  {author} {\bibinfo {author} {\bibfnamefont {S.}~\bibnamefont
  {Pirandola}},\ }\href {https://doi.org/10.1103/PhysRevLett.106.090504}
  {\bibfield  {journal} {\bibinfo  {journal} {Phys. Rev. Lett.}\ }\textbf
  {\bibinfo {volume} {106}},\ \bibinfo {pages} {090504} (\bibinfo {year}
  {2011})}\BibitemShut {NoStop}%
\bibitem [{\citenamefont {D'{A}riano}\ \emph {et~al.}(2007)\citenamefont
  {D'{A}riano}, \citenamefont {Kretschmann}, \citenamefont {Schlingemann},\
  and\ \citenamefont {Werner}}]{Dar-Kre-Sch-Wer-2007}%
  \BibitemOpen
  \bibfield  {author} {\bibinfo {author} {\bibfnamefont {G.~M.}\ \bibnamefont
  {D'{A}riano}}, \bibinfo {author} {\bibfnamefont {D.}~\bibnamefont
  {Kretschmann}}, \bibinfo {author} {\bibfnamefont {D.}~\bibnamefont
  {Schlingemann}}, \ and\ \bibinfo {author} {\bibfnamefont {R.~F.}\
  \bibnamefont {Werner}},\ }\href {https://doi.org/10.1103/PhysRevA.76.032328}
  {\bibfield  {journal} {\bibinfo  {journal} {Phys. Rev. A}\ }\textbf {\bibinfo
  {volume} {76}},\ \bibinfo {pages} {032328} (\bibinfo {year}
  {2007})}\BibitemShut {NoStop}%
\bibitem [{\citenamefont {Gutoski}\ and\ \citenamefont
  {Watrous}(2007)}]{Gut-Wat-2007}%
  \BibitemOpen
  \bibfield  {author} {\bibinfo {author} {\bibfnamefont {G.}~\bibnamefont
  {Gutoski}}\ and\ \bibinfo {author} {\bibfnamefont {J.}~\bibnamefont
  {Watrous}},\ }in\ \href {https://doi.org/10.1145/1250790.1250873} {\emph
  {\bibinfo {booktitle} {Proceedings of the 39th annual ACM symposium on Theory
  of computing}}}\ (\bibinfo {year} {2007})\ pp.\ \bibinfo {pages}
  {565--574}\BibitemShut {NoStop}%
\bibitem [{\citenamefont {Zhuang}\ and\ \citenamefont
  {Pirandola}(2020{\natexlab{a}})}]{Zhu-Pir-2020}%
  \BibitemOpen
  \bibfield  {author} {\bibinfo {author} {\bibfnamefont {Q.}~\bibnamefont
  {Zhuang}}\ and\ \bibinfo {author} {\bibfnamefont {S.}~\bibnamefont
  {Pirandola}},\ }\href {https://doi.org/10.1103/PhysRevLett.125.080505}
  {\bibfield  {journal} {\bibinfo  {journal} {Phys. Rev. Lett.}\ }\textbf
  {\bibinfo {volume} {125}},\ \bibinfo {pages} {080505} (\bibinfo {year}
  {2020}{\natexlab{a}})}\BibitemShut {NoStop}%
\bibitem [{\citenamefont {Ishizaka}\ and\ \citenamefont
  {Hiroshima}(2008)}]{Ish-Hir-2008}%
  \BibitemOpen
  \bibfield  {author} {\bibinfo {author} {\bibfnamefont {S.}~\bibnamefont
  {Ishizaka}}\ and\ \bibinfo {author} {\bibfnamefont {T.}~\bibnamefont
  {Hiroshima}},\ }\href {https://doi.org/10.1103/PhysRevLett.101.240501}
  {\bibfield  {journal} {\bibinfo  {journal} {Phys. Rev. Lett.}\ }\textbf
  {\bibinfo {volume} {101}},\ \bibinfo {pages} {240501} (\bibinfo {year}
  {2008})}\BibitemShut {NoStop}%
\bibitem [{\citenamefont {Ishizaka}\ and\ \citenamefont
  {Hiroshima}(2009)}]{Ish-Hir-2009}%
  \BibitemOpen
  \bibfield  {author} {\bibinfo {author} {\bibfnamefont {S.}~\bibnamefont
  {Ishizaka}}\ and\ \bibinfo {author} {\bibfnamefont {T.}~\bibnamefont
  {Hiroshima}},\ }\href {https://doi.org/10.1103/PhysRevA.79.042306} {\bibfield
   {journal} {\bibinfo  {journal} {Phys. Rev. A}\ }\textbf {\bibinfo {volume}
  {79}},\ \bibinfo {pages} {042306} (\bibinfo {year} {2009})}\BibitemShut
  {NoStop}%
\bibitem [{\citenamefont {Jen{\v{c}}ov{\'a}}\ and\ \citenamefont
  {Pl{\'a}vala}(2016)}]{Jen-Pla-2016}%
  \BibitemOpen
  \bibfield  {author} {\bibinfo {author} {\bibfnamefont {A.}~\bibnamefont
  {Jen{\v{c}}ov{\'a}}}\ and\ \bibinfo {author} {\bibfnamefont {M.}~\bibnamefont
  {Pl{\'a}vala}},\ }\href {https://doi.org/10.1063/1.4972286} {\bibfield
  {journal} {\bibinfo  {journal} {J. Math. Phys.}\ }\textbf {\bibinfo {volume}
  {57}},\ \bibinfo {pages} {122203} (\bibinfo {year} {2016})}\BibitemShut
  {NoStop}%
\bibitem [{\citenamefont {Hayashi}(2009)}]{Hay-2009}%
  \BibitemOpen
  \bibfield  {author} {\bibinfo {author} {\bibfnamefont {M.}~\bibnamefont
  {Hayashi}},\ }\href {https://doi.org/10.1109/TIT.2009.2023726} {\bibfield
  {journal} {\bibinfo  {journal} {IEEE Trans. Inf. Theory}\ }\textbf {\bibinfo
  {volume} {55}},\ \bibinfo {pages} {3807} (\bibinfo {year}
  {2009})}\BibitemShut {NoStop}%
\bibitem [{\citenamefont {Duan}\ \emph {et~al.}(2016)\citenamefont {Duan},
  \citenamefont {Guo}, \citenamefont {Li},\ and\ \citenamefont
  {Li}}]{Dua-Guo-Li-Li-2016}%
  \BibitemOpen
  \bibfield  {author} {\bibinfo {author} {\bibfnamefont {R.}~\bibnamefont
  {Duan}}, \bibinfo {author} {\bibfnamefont {C.}~\bibnamefont {Guo}}, \bibinfo
  {author} {\bibfnamefont {C.-K.}\ \bibnamefont {Li}}, \ and\ \bibinfo {author}
  {\bibfnamefont {Y.}~\bibnamefont {Li}},\ }in\ \href
  {https://doi.org/10.1109/ISIT.2016.7541701} {\emph {\bibinfo {booktitle}
  {Proc. IEEE Int. Symp. Inf. Theory (ISIT)}}}\ (\bibinfo {organization}
  {IEEE},\ \bibinfo {year} {2016})\ pp.\ \bibinfo {pages}
  {2259--2263}\BibitemShut {NoStop}%
\bibitem [{\citenamefont {Pirandola}\ \emph {et~al.}(2018)\citenamefont
  {Pirandola}, \citenamefont {Bardhan}, \citenamefont {Gehring}, \citenamefont
  {Weedbrook},\ and\ \citenamefont {Lloyd}}]{Pir-Bar-Geh-Wee-2018}%
  \BibitemOpen
  \bibfield  {author} {\bibinfo {author} {\bibfnamefont {S.}~\bibnamefont
  {Pirandola}}, \bibinfo {author} {\bibfnamefont {B.~R.}\ \bibnamefont
  {Bardhan}}, \bibinfo {author} {\bibfnamefont {T.}~\bibnamefont {Gehring}},
  \bibinfo {author} {\bibfnamefont {C.}~\bibnamefont {Weedbrook}}, \ and\
  \bibinfo {author} {\bibfnamefont {S.}~\bibnamefont {Lloyd}},\ }\href
  {https://doi.org/10.1038/s41566-018-0301-6} {\bibfield  {journal} {\bibinfo
  {journal} {Nat. Photonics}\ }\textbf {\bibinfo {volume} {12}},\ \bibinfo
  {pages} {724} (\bibinfo {year} {2018})}\BibitemShut {NoStop}%
\bibitem [{\citenamefont {Katariya}\ and\ \citenamefont
  {Wilde}(2020)}]{Kat-Wil-2020}%
  \BibitemOpen
  \bibfield  {author} {\bibinfo {author} {\bibfnamefont {V.}~\bibnamefont
  {Katariya}}\ and\ \bibinfo {author} {\bibfnamefont {M.~M.}\ \bibnamefont
  {Wilde}},\ }\href {https://arxiv.org/abs/2001.05376} {\bibfield  {journal}
  {\bibinfo  {journal} {arXiv preprint arXiv:2001.05376}\ } (\bibinfo {year}
  {2020})}\BibitemShut {NoStop}%
\bibitem [{\citenamefont {Pucha{\l}a}\ \emph {et~al.}(2021)\citenamefont
  {Pucha{\l}a}, \citenamefont {Pawela}, \citenamefont {Krawiec}, \citenamefont
  {Kukulski},\ and\ \citenamefont {Oszmaniec}}]{Puc-Paw-Kra-Kuk-2021}%
  \BibitemOpen
  \bibfield  {author} {\bibinfo {author} {\bibfnamefont {Z.}~\bibnamefont
  {Pucha{\l}a}}, \bibinfo {author} {\bibfnamefont {{\L}.}~\bibnamefont
  {Pawela}}, \bibinfo {author} {\bibfnamefont {A.}~\bibnamefont {Krawiec}},
  \bibinfo {author} {\bibfnamefont {R.}~\bibnamefont {Kukulski}}, \ and\
  \bibinfo {author} {\bibfnamefont {M.}~\bibnamefont {Oszmaniec}},\ }\href
  {https://doi.org/10.22331/q-2021-04-06-425} {\bibfield  {journal} {\bibinfo
  {journal} {Quantum}\ }\textbf {\bibinfo {volume} {5}},\ \bibinfo {pages}
  {425} (\bibinfo {year} {2021})}\BibitemShut {NoStop}%
\bibitem [{\citenamefont {Harrow}\ \emph {et~al.}(2010)\citenamefont {Harrow},
  \citenamefont {Hassidim}, \citenamefont {Leung},\ and\ \citenamefont
  {Watrous}}]{Har-Has-Leu-Wat-2010}%
  \BibitemOpen
  \bibfield  {author} {\bibinfo {author} {\bibfnamefont {A.~W.}\ \bibnamefont
  {Harrow}}, \bibinfo {author} {\bibfnamefont {A.}~\bibnamefont {Hassidim}},
  \bibinfo {author} {\bibfnamefont {D.~W.}\ \bibnamefont {Leung}}, \ and\
  \bibinfo {author} {\bibfnamefont {J.}~\bibnamefont {Watrous}},\ }\href
  {https://doi.org/10.1103/PhysRevA.81.032339} {\bibfield  {journal} {\bibinfo
  {journal} {Phys. Rev. A}\ }\textbf {\bibinfo {volume} {81}},\ \bibinfo
  {pages} {032339} (\bibinfo {year} {2010})}\BibitemShut {NoStop}%
\bibitem [{\citenamefont {Bondurant}(1993)}]{Bon-1993}%
  \BibitemOpen
  \bibfield  {author} {\bibinfo {author} {\bibfnamefont {R.~S.}\ \bibnamefont
  {Bondurant}},\ }\href {https://doi.org/10.1364/OL.18.001896} {\bibfield
  {journal} {\bibinfo  {journal} {Optics Letters}\ }\textbf {\bibinfo {volume}
  {18}},\ \bibinfo {pages} {1896} (\bibinfo {year} {1993})}\BibitemShut
  {NoStop}%
\bibitem [{\citenamefont {Assalini}\ \emph {et~al.}(2011)\citenamefont
  {Assalini}, \citenamefont {Dalla~Pozza},\ and\ \citenamefont
  {Pierobon}}]{Ass-Poz-Pie-2011}%
  \BibitemOpen
  \bibfield  {author} {\bibinfo {author} {\bibfnamefont {A.}~\bibnamefont
  {Assalini}}, \bibinfo {author} {\bibfnamefont {N.}~\bibnamefont
  {Dalla~Pozza}}, \ and\ \bibinfo {author} {\bibfnamefont {G.}~\bibnamefont
  {Pierobon}},\ }\href {https://doi.org/10.1103/PhysRevA.84.022342} {\bibfield
  {journal} {\bibinfo  {journal} {Phys. Rev. A}\ }\textbf {\bibinfo {volume}
  {84}},\ \bibinfo {pages} {022342} (\bibinfo {year} {2011})}\BibitemShut
  {NoStop}%
\bibitem [{\citenamefont {Becerra}\ \emph {et~al.}(2013)\citenamefont
  {Becerra}, \citenamefont {Fan}, \citenamefont {Baumgartner}, \citenamefont
  {Goldhar}, \citenamefont {Kosloski},\ and\ \citenamefont
  {Migdall}}]{Bec-Fan-Bau-Gol-2013}%
  \BibitemOpen
  \bibfield  {author} {\bibinfo {author} {\bibfnamefont {F.}~\bibnamefont
  {Becerra}}, \bibinfo {author} {\bibfnamefont {J.}~\bibnamefont {Fan}},
  \bibinfo {author} {\bibfnamefont {G.}~\bibnamefont {Baumgartner}}, \bibinfo
  {author} {\bibfnamefont {J.}~\bibnamefont {Goldhar}}, \bibinfo {author}
  {\bibfnamefont {J.}~\bibnamefont {Kosloski}}, \ and\ \bibinfo {author}
  {\bibfnamefont {A.}~\bibnamefont {Migdall}},\ }\href
  {https://doi.org/10.1038/nphoton.2012.316} {\bibfield  {journal} {\bibinfo
  {journal} {Nat. Photonics}\ }\textbf {\bibinfo {volume} {7}},\ \bibinfo
  {pages} {147} (\bibinfo {year} {2013})}\BibitemShut {NoStop}%
\bibitem [{\citenamefont {Flatt}\ \emph {et~al.}(2019)\citenamefont {Flatt},
  \citenamefont {Barnett},\ and\ \citenamefont {Croke}}]{Fla-Bar-Cro-2019}%
  \BibitemOpen
  \bibfield  {author} {\bibinfo {author} {\bibfnamefont {K.}~\bibnamefont
  {Flatt}}, \bibinfo {author} {\bibfnamefont {S.~M.}\ \bibnamefont {Barnett}},
  \ and\ \bibinfo {author} {\bibfnamefont {S.}~\bibnamefont {Croke}},\ }\href
  {https://doi.org/10.1103/PhysRevA.100.032122} {\bibfield  {journal} {\bibinfo
   {journal} {Phys. Rev. A}\ }\textbf {\bibinfo {volume} {100}},\ \bibinfo
  {pages} {032122} (\bibinfo {year} {2019})}\BibitemShut {NoStop}%
\bibitem [{\citenamefont {Dolinar}(1976)}]{Dol-1976}%
  \BibitemOpen
  \bibfield  {author} {\bibinfo {author} {\bibfnamefont {S.~J.}\ \bibnamefont
  {Dolinar}},\ }\href@noop {} {\emph {\bibinfo {title} {A class of optical
  receivers using optical feedback}}},\ Vol.\ \bibinfo {volume} {111}\
  (\bibinfo  {publisher} {Ph.D. dissertation, Massachusetts Institute of
  Technology},\ \bibinfo {address} {Cambridge, MA},\ \bibinfo {year}
  {1976})\BibitemShut {NoStop}%
\bibitem [{\citenamefont {Brody}\ and\ \citenamefont
  {Meister}(1996)}]{Bro-Mei-1996}%
  \BibitemOpen
  \bibfield  {author} {\bibinfo {author} {\bibfnamefont {D.}~\bibnamefont
  {Brody}}\ and\ \bibinfo {author} {\bibfnamefont {B.}~\bibnamefont
  {Meister}},\ }\href {https://doi.org/10.1103/PhysRevLett.76.1} {\bibfield
  {journal} {\bibinfo  {journal} {Phys. Rev. Lett.}\ }\textbf {\bibinfo
  {volume} {76}},\ \bibinfo {pages} {1} (\bibinfo {year} {1996})}\BibitemShut
  {NoStop}%
\bibitem [{\citenamefont {Acin}\ \emph {et~al.}(2005)\citenamefont {Acin},
  \citenamefont {Bagan}, \citenamefont {Baig}, \citenamefont {Masanes},\ and\
  \citenamefont {Mu{\~n}oz-Tapia}}]{Aci-Bag-Bai-Mas-2005}%
  \BibitemOpen
  \bibfield  {author} {\bibinfo {author} {\bibfnamefont {A.}~\bibnamefont
  {Acin}}, \bibinfo {author} {\bibfnamefont {E.}~\bibnamefont {Bagan}},
  \bibinfo {author} {\bibfnamefont {M.}~\bibnamefont {Baig}}, \bibinfo {author}
  {\bibfnamefont {L.}~\bibnamefont {Masanes}}, \ and\ \bibinfo {author}
  {\bibfnamefont {R.}~\bibnamefont {Mu{\~n}oz-Tapia}},\ }\href
  {https://doi.org/10.1103/PhysRevA.71.032338} {\bibfield  {journal} {\bibinfo
  {journal} {Phys. Rev. A}\ }\textbf {\bibinfo {volume} {71}},\ \bibinfo
  {pages} {032338} (\bibinfo {year} {2005})}\BibitemShut {NoStop}%
\bibitem [{\citenamefont {Ludwig}(1987)}]{Lud-1985}%
  \BibitemOpen
  \bibfield  {author} {\bibinfo {author} {\bibfnamefont {G.}~\bibnamefont
  {Ludwig}},\ }\href@noop {} {\emph {\bibinfo {title} {An axiomatic basis of
  quantum mechanics. vols. I and II}}}\ (\bibinfo  {publisher} {Springer},\
  \bibinfo {year} {1985 and 1987})\BibitemShut {NoStop}%
\bibitem [{\citenamefont {Hartk{\"a}mper}\ and\ \citenamefont
  {Neumann}(1974)}]{Har-Neu-1974}%
  \BibitemOpen
  \bibinfo {editor} {\bibfnamefont {A.}~\bibnamefont {Hartk{\"a}mper}}\ and\
  \bibinfo {editor} {\bibfnamefont {H.}~\bibnamefont {Neumann}},\ eds.,\ \href
  {https://doi.org/10.1007/3-540-06725-6} {\emph {\bibinfo {title} {Foundations
  of quantum mechanics and ordered linear spaces}}}\ (\bibinfo  {publisher}
  {Springer},\ \bibinfo {address} {New York},\ \bibinfo {year}
  {1974})\BibitemShut {NoStop}%
\bibitem [{\citenamefont {Barrett}(2007)}]{Bar-2007}%
  \BibitemOpen
  \bibfield  {author} {\bibinfo {author} {\bibfnamefont {J.}~\bibnamefont
  {Barrett}},\ }\href {https://doi.org/10.1103/PhysRevA.75.032304} {\bibfield
  {journal} {\bibinfo  {journal} {Phys. Rev. A}\ }\textbf {\bibinfo {volume}
  {75}},\ \bibinfo {pages} {032304} (\bibinfo {year} {2007})}\BibitemShut
  {NoStop}%
\bibitem [{\citenamefont {Chiribella}\ \emph {et~al.}(2010)\citenamefont
  {Chiribella}, \citenamefont {D'{A}riano},\ and\ \citenamefont
  {Perinotti}}]{Chi-Dar-Per-2010}%
  \BibitemOpen
  \bibfield  {author} {\bibinfo {author} {\bibfnamefont {G.}~\bibnamefont
  {Chiribella}}, \bibinfo {author} {\bibfnamefont {G.~M.}\ \bibnamefont
  {D'{A}riano}}, \ and\ \bibinfo {author} {\bibfnamefont {P.}~\bibnamefont
  {Perinotti}},\ }\href {https://doi.org/10.1103/PhysRevA.81.062348} {\bibfield
   {journal} {\bibinfo  {journal} {Phys. Rev. A}\ }\textbf {\bibinfo {volume}
  {81}},\ \bibinfo {pages} {062348} (\bibinfo {year} {2010})}\BibitemShut
  {NoStop}%
\bibitem [{\citenamefont {Janotta}\ and\ \citenamefont
  {Lal}(2013)}]{Jan-Lal-2013}%
  \BibitemOpen
  \bibfield  {author} {\bibinfo {author} {\bibfnamefont {P.}~\bibnamefont
  {Janotta}}\ and\ \bibinfo {author} {\bibfnamefont {R.}~\bibnamefont {Lal}},\
  }\href {https://doi.org/10.1103/PhysRevA.87.052131} {\bibfield  {journal}
  {\bibinfo  {journal} {Phys. Rev. A}\ }\textbf {\bibinfo {volume} {87}},\
  \bibinfo {pages} {052131} (\bibinfo {year} {2013})}\BibitemShut {NoStop}%
\bibitem [{\citenamefont {Chiribella}\ \emph
  {et~al.}(2008{\natexlab{c}})\citenamefont {Chiribella}, \citenamefont
  {D'{A}riano},\ and\ \citenamefont {Perinotti}}]{Chi-Dar-Per-2008-memory}%
  \BibitemOpen
  \bibfield  {author} {\bibinfo {author} {\bibfnamefont {G.}~\bibnamefont
  {Chiribella}}, \bibinfo {author} {\bibfnamefont {G.~M.}\ \bibnamefont
  {D'{A}riano}}, \ and\ \bibinfo {author} {\bibfnamefont {P.}~\bibnamefont
  {Perinotti}},\ }\href {https://doi.org/10.1103/PhysRevLett.101.180501}
  {\bibfield  {journal} {\bibinfo  {journal} {Phys. Rev. Lett.}\ }\textbf
  {\bibinfo {volume} {101}},\ \bibinfo {pages} {180501} (\bibinfo {year}
  {2008}{\natexlab{c}})}\BibitemShut {NoStop}%
\bibitem [{\citenamefont {Chiribella}(2012)}]{Chi-2012}%
  \BibitemOpen
  \bibfield  {author} {\bibinfo {author} {\bibfnamefont {G.}~\bibnamefont
  {Chiribella}},\ }\href {https://doi.org/10.1088/1367-2630/14/12/125008}
  {\bibfield  {journal} {\bibinfo  {journal} {New J. Phys.}\ }\textbf {\bibinfo
  {volume} {14}},\ \bibinfo {pages} {125008} (\bibinfo {year}
  {2012})}\BibitemShut {NoStop}%
\bibitem [{Note1()}]{Note1}%
  \BibitemOpen
  \bibinfo {note} {Let $(s_t^\star ,X_t^\star )$ and $(s_{t-1}^\star
  ,X_{t-1}^\star )$ be, respectively, the optimal solutions to Eq.~\protect
  \eqref {prob:s} and that with $t$ replaced by $t-1$; then, we can easily
  verify that $(s_{t-1}^\star s_t^\star ,X_{t-1}^\star ,X_t^\star )$ is a
  feasible solution to Eq.~\protect \eqref {prob:s2}, which yields
  $s_{t-1}^\star s_t^\star \ge s_{t,2}^\star $. Thus, $\protect \overline {P_2}
  \le \protect \overline {P_1}$ holds. By the same discussion as $P\le \protect
  \overline {P_1}$, we obtain $P\le \protect \overline {P_2}$.}\BibitemShut
  {Stop}%
\bibitem [{\citenamefont {Choi}(1975)}]{Cho-1975}%
  \BibitemOpen
  \bibfield  {author} {\bibinfo {author} {\bibfnamefont {M.-D.}\ \bibnamefont
  {Choi}},\ }\href {https://doi.org/10.1016/0024-3795(75)90075-0} {\bibfield
  {journal} {\bibinfo  {journal} {Lin. Alg. Appl.}\ }\textbf {\bibinfo {volume}
  {10}},\ \bibinfo {pages} {285} (\bibinfo {year} {1975})}\BibitemShut
  {NoStop}%
\bibitem [{\citenamefont {Jamio{\l}kowski}(1972)}]{Jam-1972}%
  \BibitemOpen
  \bibfield  {author} {\bibinfo {author} {\bibfnamefont {A.}~\bibnamefont
  {Jamio{\l}kowski}},\ }\href {https://doi.org/10.1016/0034-4877(72)90011-0}
  {\bibfield  {journal} {\bibinfo  {journal} {Rep. Math. Phys.}\ }\textbf
  {\bibinfo {volume} {3}},\ \bibinfo {pages} {275} (\bibinfo {year}
  {1972})}\BibitemShut {NoStop}%
\bibitem [{\citenamefont {Nakahira}\ and\ \citenamefont
  {Kato}(2021)}]{Nak-Kat-2021-general}%
  \BibitemOpen
  \bibfield  {author} {\bibinfo {author} {\bibfnamefont {K.}~\bibnamefont
  {Nakahira}}\ and\ \bibinfo {author} {\bibfnamefont {K.}~\bibnamefont
  {Kato}},\ }\href {https://arxiv.org/abs/2104.09759} {\bibfield  {journal}
  {\bibinfo  {journal} {arXiv preprint arXiv:2104.09759}\ } (\bibinfo {year}
  {2021})}\BibitemShut {NoStop}%
\bibitem [{SM()}]{SM}%
  \BibitemOpen
  \href@noop {} {}\bibinfo {note} {See Supplemental Material, which includes
  Ref.~\cite{Jeo-Shi-2019}, for additional information about the proposed upper
  and lower bounds, numerical simulations, and computational
  complexity.}\BibitemShut {Stop}%
\bibitem [{\citenamefont {Lloyd}(2008)}]{Llo-2008}%
  \BibitemOpen
  \bibfield  {author} {\bibinfo {author} {\bibfnamefont {S.}~\bibnamefont
  {Lloyd}},\ }\href {https://doi.org/10.1126/science.1160627} {\bibfield
  {journal} {\bibinfo  {journal} {Science}\ }\textbf {\bibinfo {volume}
  {321}},\ \bibinfo {pages} {1463} (\bibinfo {year} {2008})}\BibitemShut
  {NoStop}%
\bibitem [{\citenamefont {Tan}\ \emph {et~al.}(2008)\citenamefont {Tan},
  \citenamefont {Erkmen}, \citenamefont {Giovannetti}, \citenamefont {Guha},
  \citenamefont {Lloyd}, \citenamefont {Maccone}, \citenamefont {Pirandola},\
  and\ \citenamefont {Shapiro}}]{Tan-Erk-Gio-Guh-2008}%
  \BibitemOpen
  \bibfield  {author} {\bibinfo {author} {\bibfnamefont {S.-H.}\ \bibnamefont
  {Tan}}, \bibinfo {author} {\bibfnamefont {B.~I.}\ \bibnamefont {Erkmen}},
  \bibinfo {author} {\bibfnamefont {V.}~\bibnamefont {Giovannetti}}, \bibinfo
  {author} {\bibfnamefont {S.}~\bibnamefont {Guha}}, \bibinfo {author}
  {\bibfnamefont {S.}~\bibnamefont {Lloyd}}, \bibinfo {author} {\bibfnamefont
  {L.}~\bibnamefont {Maccone}}, \bibinfo {author} {\bibfnamefont
  {S.}~\bibnamefont {Pirandola}}, \ and\ \bibinfo {author} {\bibfnamefont
  {J.~H.}\ \bibnamefont {Shapiro}},\ }\href
  {https://doi.org/10.1103/PhysRevLett.101.253601} {\bibfield  {journal}
  {\bibinfo  {journal} {Phys. Rev. Lett.}\ }\textbf {\bibinfo {volume} {101}},\
  \bibinfo {pages} {253601} (\bibinfo {year} {2008})}\BibitemShut {NoStop}%
\bibitem [{\citenamefont {Holevo}(1978)}]{Hol-1978}%
  \BibitemOpen
  \bibfield  {author} {\bibinfo {author} {\bibfnamefont {A.~S.}\ \bibnamefont
  {Holevo}},\ }\href {https://doi.org/10.1137/1123048} {\bibfield  {journal}
  {\bibinfo  {journal} {Teor. Veroyatnost. i Primenen.}\ }\textbf {\bibinfo
  {volume} {23}},\ \bibinfo {pages} {429} (\bibinfo {year} {1978})}\BibitemShut
  {NoStop}%
\bibitem [{\citenamefont {Hausladen}\ and\ \citenamefont
  {Wootters}(1994)}]{Hau-Woo-1994}%
  \BibitemOpen
  \bibfield  {author} {\bibinfo {author} {\bibfnamefont {P.}~\bibnamefont
  {Hausladen}}\ and\ \bibinfo {author} {\bibfnamefont {W.~K.}\ \bibnamefont
  {Wootters}},\ }\href {https://doi.org/10.1080/09500349414552221} {\bibfield
  {journal} {\bibinfo  {journal} {J. Mod. Opt.}\ }\textbf {\bibinfo {volume}
  {41}},\ \bibinfo {pages} {2385} (\bibinfo {year} {1994})}\BibitemShut
  {NoStop}%
\bibitem [{\citenamefont {Zhuang}\ and\ \citenamefont
  {Pirandola}(2020{\natexlab{b}})}]{Zhu-Pir-2020-entangle}%
  \BibitemOpen
  \bibfield  {author} {\bibinfo {author} {\bibfnamefont {Q.}~\bibnamefont
  {Zhuang}}\ and\ \bibinfo {author} {\bibfnamefont {S.}~\bibnamefont
  {Pirandola}},\ }\href {https://doi.org/10.1038/s42005-020-0369-4} {\bibfield
  {journal} {\bibinfo  {journal} {Commun. Phys.}\ }\textbf {\bibinfo {volume}
  {3}},\ \bibinfo {pages} {1} (\bibinfo {year}
  {2020}{\natexlab{b}})}\BibitemShut {NoStop}%
\bibitem [{\citenamefont {Borchers}(1999)}]{Bor-1999}%
  \BibitemOpen
  \bibfield  {author} {\bibinfo {author} {\bibfnamefont {B.}~\bibnamefont
  {Borchers}},\ }\href {https://doi.org/10.1080/10556789908805765} {\bibfield
  {journal} {\bibinfo  {journal} {Optimization Methods and Software}\ }\textbf
  {\bibinfo {volume} {11}},\ \bibinfo {pages} {613} (\bibinfo {year}
  {1999})}\BibitemShut {NoStop}%
\bibitem [{Note2()}]{Note2}%
  \BibitemOpen
  \bibinfo {note} {From Eq.~(7) of Ref.~\cite {Zhu-Pir-2020}, we have $\protect
  \overline {P}_\protect \mathrm {conv} \ge 1 - (M-1)/2M =
  (M+1)/2M$.}\BibitemShut {Stop}%
\bibitem [{Note3()}]{Note3}%
  \BibitemOpen
  \bibinfo {note} {Let $P^\star _1$ be the ultimate success probability in the
  case of $T = 1$; then, $\protect \overline {P_1} = M^{T-1} P^\star _1{}^T$
  holds from $\protect \overline {P_1} = s^\star {}^T/M$ and $P^\star _1 =
  s^\star /M$, where $s^\star $ is the optimal value of Problem~\protect \eqref
  {prob:s}. It follows from $\protect \overline {P}_\protect \mathrm {conv} \ge
  (M+1)/2M$ that $\protect \overline {P_1}$ is tighter than $\protect \overline
  {P}_\protect \mathrm {conv}$ whenever $P^\star _1 <
  [(M+1)/2]^{1/T}/M$.}\BibitemShut {Stop}%
\bibitem [{Note4()}]{Note4}%
  \BibitemOpen
  \bibinfo {note} {We did this numerical experiment on a PC with 16~GB memory,
  in which case neither $\protect \overline {P_1}$ for $M \ge 4$ nor
  $P_\protect \mathrm {PGM}$ for $TM \ge 9$ can be computed due to memory
  limitations.}\BibitemShut {Stop}%
\bibitem [{\citenamefont {Jeong}\ and\ \citenamefont
  {Shin}(2019)}]{Jeo-Shi-2019}%
  \BibitemOpen
  \bibfield  {author} {\bibinfo {author} {\bibfnamefont {Y.}~\bibnamefont
  {Jeong}}\ and\ \bibinfo {author} {\bibfnamefont {H.}~\bibnamefont {Shin}},\
  }\href {https://doi.org/10.1038/s41598-019-40652-0} {\bibfield  {journal}
  {\bibinfo  {journal} {Sci. Rep.}\ }\textbf {\bibinfo {volume} {9}},\ \bibinfo
  {pages} {4035} (\bibinfo {year} {2019})}\BibitemShut {NoStop}%
\end{thebibliography}

\begin{thebibliography}{6}%
\makeatletter
\providecommand \@ifxundefined [1]{%
 \@ifx{#1\undefined}
}%
\providecommand \@ifnum [1]{%
 \ifnum #1\expandafter \@firstoftwo
 \else \expandafter \@secondoftwo
 \fi
}%
\providecommand \@ifx [1]{%
 \ifx #1\expandafter \@firstoftwo
 \else \expandafter \@secondoftwo
 \fi
}%
\providecommand \natexlab [1]{#1}%
\providecommand \enquote  [1]{``#1''}%
\providecommand \bibnamefont  [1]{#1}%
\providecommand \bibfnamefont [1]{#1}%
\providecommand \citenamefont [1]{#1}%
\providecommand \href@noop [0]{\@secondoftwo}%
\providecommand \href [0]{\begingroup \@sanitize@url \@href}%
\providecommand \@href[1]{\@@startlink{#1}\@@href}%
\providecommand \@@href[1]{\endgroup#1\@@endlink}%
\providecommand \@sanitize@url [0]{\catcode `\\12\catcode `\$12\catcode
  `\&12\catcode `\#12\catcode `\^12\catcode `\_12\catcode `\%12\relax}%
\providecommand \@@startlink[1]{}%
\providecommand \@@endlink[0]{}%
\providecommand \url  [0]{\begingroup\@sanitize@url \@url }%
\providecommand \@url [1]{\endgroup\@href {#1}{\urlprefix }}%
\providecommand \urlprefix  [0]{URL }%
\providecommand \Eprint [0]{\href }%
\providecommand \doibase [0]{http://dx.doi.org/}%
\providecommand \selectlanguage [0]{\@gobble}%
\providecommand \bibinfo  [0]{\@secondoftwo}%
\providecommand \bibfield  [0]{\@secondoftwo}%
\providecommand \translation [1]{[#1]}%
\providecommand \BibitemOpen [0]{}%
\providecommand \bibitemStop [0]{}%
\providecommand \bibitemNoStop [0]{.\EOS\space}%
\providecommand \EOS [0]{\spacefactor3000\relax}%
\providecommand \BibitemShut  [1]{\csname bibitem#1\endcsname}%
\let\auto@bib@innerbib\@empty
%</preamble>
\bibitem [{\citenamefont {Chiribella}\ \emph {et~al.}(2008)\citenamefont
  {Chiribella}, \citenamefont {D'{A}riano},\ and\ \citenamefont
  {Perinotti}}]{Chi-Dar-Per-2008-sm}%
  \BibitemOpen
  \bibfield  {author} {\bibinfo {author} {\bibfnamefont {G.}~\bibnamefont
  {Chiribella}}, \bibinfo {author} {\bibfnamefont {G.~M.}\ \bibnamefont
  {D'{A}riano}}, \ and\ \bibinfo {author} {\bibfnamefont {P.}~\bibnamefont
  {Perinotti}},\ }\href {https://doi.org/10.1103/PhysRevLett.101.060401}
  {\bibfield  {journal} {\bibinfo  {journal} {Phys. Rev. Lett.}\ }\textbf
  {\bibinfo {volume} {101}},\ \bibinfo {pages} {060401} (\bibinfo {year}
  {2008})}\BibitemShut {NoStop}%
\bibitem [{\citenamefont {Chiribella}(2012)}]{Chi-2012-sm}%
  \BibitemOpen
  \bibfield  {author} {\bibinfo {author} {\bibfnamefont {G.}~\bibnamefont
  {Chiribella}},\ }\href {https://doi.org/10.1088/1367-2630/14/12/125008}
  {\bibfield  {journal} {\bibinfo  {journal} {New J. Phys.}\ }\textbf {\bibinfo
  {volume} {14}},\ \bibinfo {pages} {125008} (\bibinfo {year}
  {2012})}\BibitemShut {NoStop}%
\bibitem [{\citenamefont {Zhuang}\ and\ \citenamefont
  {Pirandola}(2020)}]{Zhu-Pir-2020-sm}%
  \BibitemOpen
  \bibfield  {author} {\bibinfo {author} {\bibfnamefont {Q.}~\bibnamefont
  {Zhuang}}\ and\ \bibinfo {author} {\bibfnamefont {S.}~\bibnamefont
  {Pirandola}},\ }\href {https://doi.org/10.1103/PhysRevLett.125.080505}
  {\bibfield  {journal} {\bibinfo  {journal} {Phys. Rev. Lett.}\ }\textbf
  {\bibinfo {volume} {125}},\ \bibinfo {pages} {080505} (\bibinfo {year}
  {2020})}\BibitemShut {NoStop}%
\bibitem [{\citenamefont {Jen{\v{c}}ov{\'a}}\ and\ \citenamefont
  {Pl{\'a}vala}(2016)}]{Jen-Pla-2016-sm}%
  \BibitemOpen
  \bibfield  {author} {\bibinfo {author} {\bibfnamefont {A.}~\bibnamefont
  {Jen{\v{c}}ov{\'a}}}\ and\ \bibinfo {author} {\bibfnamefont {M.}~\bibnamefont
  {Pl{\'a}vala}},\ }\href {https://doi.org/10.1063/1.4972286} {\bibfield
  {journal} {\bibinfo  {journal} {J. Math. Phys.}\ }\textbf {\bibinfo {volume}
  {57}},\ \bibinfo {pages} {122203} (\bibinfo {year} {2016})}\BibitemShut
  {NoStop}%
\bibitem [{\citenamefont {Jeong}\ and\ \citenamefont
  {Shin}(2019)}]{Jeo-Shi-2019-sm}%
  \BibitemOpen
  \bibfield  {author} {\bibinfo {author} {\bibfnamefont {Y.}~\bibnamefont
  {Jeong}}\ and\ \bibinfo {author} {\bibfnamefont {H.}~\bibnamefont {Shin}},\
  }\href {https://doi.org/10.1038/s41598-019-40652-0} {\bibfield  {journal}
  {\bibinfo  {journal} {Sci. Rep.}\ }\textbf {\bibinfo {volume} {9}},\ \bibinfo
  {pages} {1} (\bibinfo {year} {2019})}\BibitemShut {NoStop}%
\bibitem [{\citenamefont {Borchers}(1999)}]{Bor-1999-sm}%
  \BibitemOpen
  \bibfield  {author} {\bibinfo {author} {\bibfnamefont {B.}~\bibnamefont
  {Borchers}},\ }\href {https://doi.org/10.1080/10556789908805765} {\bibfield
  {journal} {\bibinfo  {journal} {Optimization Methods and Software}\ }\textbf
  {\bibinfo {volume} {11}},\ \bibinfo {pages} {613} (\bibinfo {year}
  {1999})}\BibitemShut {NoStop}%
\end{thebibliography}
\end{document}